\documentclass[sn-mathphys-num,iicol]{sn-jnl}

\usepackage{graphicx}%
\usepackage{multirow}%
\usepackage{amsmath,amssymb,amsfonts}%
\usepackage{amsthm}%
\usepackage{mathrsfs}%
\usepackage[title]{appendix}%
\usepackage{xcolor}%
\usepackage{textcomp}%
\usepackage{manyfoot}%
\usepackage{booktabs}
\usepackage{algorithm}%
\usepackage{algorithmicx}%
\usepackage[noend]{algpseudocode}
\usepackage{listings}%
\usepackage{tabularx}

\usepackage{longtable} 
\usepackage{csvsimple}
\usepackage{lscape}
\usepackage{siunitx}
\usepackage{array}
\usepackage{url}
\usepackage{multicol}
\usepackage{enumitem}
\usepackage{amssymb}

\definecolor{newcolor}{rgb}{.8,.349,.1}
\newcolumntype{J}[1]{>{\centering\let\newline\\\arraybackslash\hspace{0pt}}m{#1}}
\newcolumntype{K}[1]{>{\let\newline\\\arraybackslash\hspace{0pt}}m{#1}}

\algnewcommand\algorithmicforeach{\textbf{for each}}
\algdef{S}[FOR]{ForEach}[1]{\algorithmicforeach\ #1\ \algorithmicdo}

\newcolumntype{A}[1]{>{\raggedright\let\newline\\\arraybackslash\hspace{0pt}}m{#1}}
\newcolumntype{R}[1]{>{\centering\let\newline\\\arraybackslash\hspace{0pt}}m{#1}}
\newcolumntype{B}[1]{>{\raggedleft\let\newline\\\arraybackslash\hspace{0pt}}m{#1}}

\theoremstyle{thmstyleone}%
\theoremstyle{thmstyletwo}%

\theoremstyle{thmstylethree}%

\def\isanonymized{0} 
\def\istitlepage{0} 

\begin{document}

\title[Improving the generalization of deep learning models in the segmentation of mammography images]{Improving the generalization of deep learning models in the segmentation of mammography images}


\if\isanonymized1
\else

\author*[1]{\fnm{Jan} \sur{Hurtado}}\email{hurtado@tecgraf.puc-rio.br}
\equalcont{These authors contributed equally to this work.}

\author[1]{\fnm{João P.} \sur{Maia}}\email{c1920354@inf.puc-rio.br}
\equalcont{These authors contributed equally to this work.}

\author[1]{\fnm{Cesar A.} \sur{Sierra-Franco}}\email{casfranco@tecgraf.puc-rio.br}
\equalcont{These authors contributed equally to this work.}

\author[1]{\fnm{Alberto} \sur{Raposo}}\email{abraposo@tecgraf.puc-rio.br}

\affil[1]{\orgdiv{Tecgraf Institute and Department of Informatics}, \orgname{Pontifical Catholic University of Rio de Janeiro}, \orgaddress{\city{Rio de Janeiro}, \state{RJ}, \country{Brazil}}}

\fi


\abstract{
Mammography stands as the main screening method for detecting breast cancer early, enhancing treatment success rates. The segmentation of landmark structures in mammography images can aid the medical assessment in the evaluation of cancer risk and the image acquisition adequacy. We introduce a series of data-centric strategies aimed at enriching the training data for deep learning-based segmentation of landmark structures. Our approach involves augmenting the training samples through annotation-guided image intensity manipulation and style transfer to achieve better generalization than standard training procedures. These augmentations are applied in a balanced manner to ensure the model learns to process a diverse range of images generated by different vendor equipments while retaining its efficacy on the original data. We present extensive numerical and visual results that demonstrate the superior generalization capabilities of our methods when compared to the standard training. For this evaluation, we consider a large dataset that includes mammography images generated by different vendor equipments. Further, we present complementary results that show both the strengths and limitations of our methods across various scenarios. The accuracy and robustness demonstrated in the experiments suggest that our method is well-suited for integration into clinical practice.
}




\maketitle

Manuscript number of words: 10270. Abstract number of words: 192. Number of figures: 13. Number of tables: 12.

\if\istitlepage1
\else
\section{Introduction}
\label{sec:introduction}

Mammography involves utilizing low-dose x-ray technology to examine the breast, making it one of the most effective screening methods in use today. Regular mammographies are vital for early detection of breast cancer, greatly enhancing the likelihood of successful treatment. While there are various types of mammography, the most prevalent method is the digital mammography, known for generating easily manipulable high-quality 2D digital images. These images comprise two primary views applied to both breasts, known as Medio-Lateral Oblique (MLO) and Cranio-Caudal (CC), which enable healthcare professionals comprehensive analysis from multiple angles. 


Accurately identifying health conditions within digital mammography images can be influenced by factors such as interpreter expertise, acquisition quality, and patient's anatomy. Therefore, providing software assistants can support image interpretation and medical assessment, alleviating the challenges posed by these factors. Several methods were proposed in the literature for this support, mostly focusing on the detection of abnormalities, the estimation of breast density, and the prediction of cancer risk. However, one of the most critical applications involves the estimation of anatomical landmark structures \cite{sierra2024towards}, such as the nipple, pectoral muscle, fibroglandular tissue, and fatty tissue. The correct identification of these structures and their extent in the image is essential for categorizing the risk of abnormalities and evaluating image acquisition adequacy.

The estimation of the pectoral muscle is essential for aiding in the assessment of potential abnormalities and the image acquisition correctness. Some deep learning-based methods were proposed to segment the pectoral muscle contour in the MLO view, which is enough for supporting image adequacy assessment \cite{rampun2019breast,soleimani2020segmentation}. For the full pectoral muscle shape segmentation, some approaches considered U-Net architectures \cite{ali2020enhancing,rubio2021multicriteria}, generative adversarial networks \cite{guo2020automatic}, attention mechanisms \cite{yu2022pemnet}, and domain-oriented augmentation operations to achieve better generalization \cite{verboom2024deep}. 

The nipple constitutes another important landmark structure, facilitating the registration of multiple views or modalities by enabling efficient region matching and anatomical measurements. Most of the proposed methods for nipple estimation are hand-crafted, involving the analysis of breast boundary \cite{yin1994computerized, mendez1996automatic, chandrasekhar1997simple, mustra2009nipple}, texture \cite{casti2013automatic}, and convergence of fibroglandular tissue \cite{zhou2004computerized, kinoshita2008radon}. Some data-driven methods were proposed to estimate the nipple position, considering random forest classifiers \cite{jiang2019radiomic} and deep neural networks \cite{lin2019nipple}.


The fibroglandular tissue represents a critical area of concern warranting specific attention during medical assessments. Depending on the patient's unique anatomy, this tissue can exhibit varying characteristics, ranging from dense formations to more dispersed patterns, with higher density correlating to increased risk. Numerous methodologies have been proposed in the existing literature for segmenting dense fibroglandular tissue regions \cite{he2015review}, encompassing both handcrafted \cite{matsubara2001automated,el2010expectation,torres2019morphological} and data-driven methods \cite{saffari2020fully,larroza2022breast,hu2022breast}.

Several methodologies aim to integrate the segmentation of various landmark structures within a unified framework. Tiryaki et al. conduct experiments employing multiple U-Net-based models to segment the pectoral muscle, dense fibroglandular tissue regions, and fatty tissues \cite{tiryaki2022deep}. In a similar way, considering these structures and incorporating the nipple, Dubrovina et al. introduce a novel deep learning-based framework for comprehensive segmentation tasks \cite{dubrovina2018computational}. By leveraging multiple deep learning models, Bou demonstrates segmentation results encompassing more intricate structures, including vessels, calcifications, and skin, among others \cite{bou2019deep}. 

In a recent study, Sierra-Franco et al. \cite{sierra2024towards} introduced a large dataset alongside deep learning experiments for the segmentation of mammography images, encompassing both MLO and CC views. The study highlights four primary structures of interest in both views: nipple, pectoral muscle, fibroglandular tissue, and fatty tissue. We propose a data-centric approach for the improvement of the generalization of the solution introduced in this work on the processing of mammography images generated by different vendor equipments.

In this paper, we propose a set of data-centric strategies to achieve better generalization on the processing of mammography images acquired using different vendor equipment. More precisely, we introduce augmentation procedures based on image intensity manipulation and style-transfer methods, incorporating samples during training that enable the model to learn from diverse hypothetical domains. We present extensive numerical and visual results on analyzing the reference method, i.e. \cite{sierra2024towards}, and highlighting the benefits of the proposed strategies. These results demonstrate the promising potential of our strategies, making them strong candidates for integration into clinical practice. 

\section{Materials and Methods}
\subsection{Datasets}
\label{sec:datasets}

In this paper, our primary focus is on utilizing MLO view digital mammography images sourced from two distinct datasets: the private dataset introduced in \cite{sierra2024towards}, and the VinDr-Mammo dataset introduced in \cite{nguyen2023vindr}. More precisely, we compose four different datasets, each one representing a different vendor of mammography equipment. These datasets are named as GE, IMS, PLANMED, and HOLOGIC, containing MLO view mammography images generated by equipments of the General Electric, IMS Giotto, Planmed Oy, and Hologic vendors, respectively. 

The main purpose of these datasets is image segmentation, and they include annotations for four major structures of interest: the nipple, pectoral muscle, fibroglandular tissue, and fatty tissue. A team of eight annotators received training from two clinical experts to identify and delineate these structures accurately using a contour drawing tool. Then, these contours, represented as polygons, are rasterized to generate a multi-class label maps representing the structures. All the left breast images are horizontally flipped to simplify the input domain. For further details about this annotation process and how the label maps are generated, please refer to \cite{sierra2024towards}. 

Trying to uniformize the input images, all the images follow the pre-processing stablished in \cite{sierra2024towards}. The mammography images are normalized using the percentiles 2 and 98 as minimum and maximum values, then equalized using Contrast Limited Adaptive Histogram Equalization (CLAHE) \cite{clahe} with kernel size being $1/8$ of the height and width of the image, and finally re-scaled to the range $[0,255]$. For the IMS and PLANMED datasets, we include additional processing due to the different image format adopted for these cases.

\begin{figure}[!t]
    \centering
    \includegraphics[width=0.75\columnwidth]{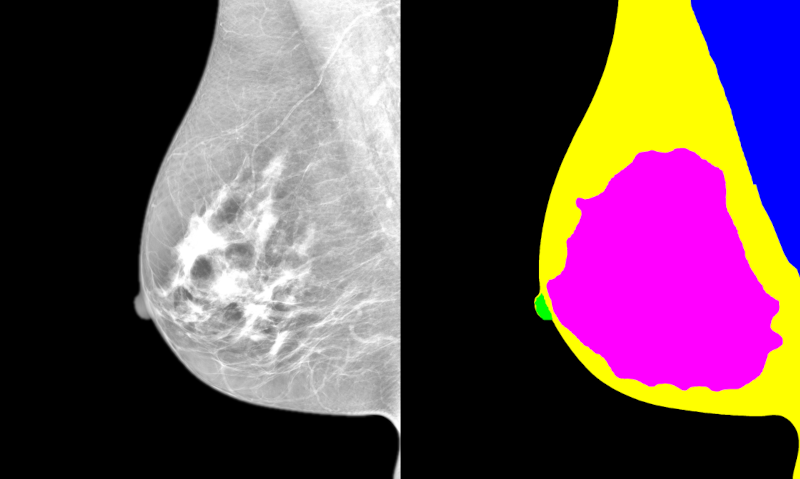}
    \caption{Pre-processed image and its corresponding label map (ground-truth annotation). The nipple is colored in green, the pectoral muscle is colored in blue, the fibroglandular tissue is colored in magenta, the fatty tissue is colored in yellow, and the background is colored in black.}
    \label{fig:ann_example}
\end{figure}

Although we present different datasets, we just consider the GE dataset for the training task due to the limited annotations available for images generated by the other vendor equipments. The IMS, PLANMED and HOLOGIC datasets are used for evaluation purposes only. We aim to train deep learning models on the GE data capable of generalizing to data from other vendors. A fully detailed specification of each dataset is presented in the following.

\textbf{GE dataset.} A collection of 5214 MLO view images was selected to construct this dataset, belonging to the acquisition of three types of GE equipments: Senographe Essential, Senograph DS, Senographe Pristina, and Senographe Crystal. All of these equipments present similar images that were fully annotated and pre-processed using the standard method explained above. The annotated samples are split into the three standard subsets considered in a conventional supervised learning pipeline: training, validation, and test. The splitting process follows a random behavior with certain balancing regarding the fibro-glandular tissue density and avoiding data leakage. This distribution results in 3450 samples for training ($\sim70\%$), 1206 samples for validation ($\sim20\%$), and 557 samples for test ($\sim10\%$). We use this dataset for both training and testing purposes. 

\textbf{IMS dataset.} This dataset comprises a collection of 52 MLO view images acquired using the GIOTTO CLASS and GIOTTO IMAGE 3DL equipment. Unlike the GE images, an additional intensity rescaling operation is included in the preprocessing pipeline for this dataset, applied prior to the previously described operations. Employing the standard window level visualization setting, we utilize the window center $c$ and window width $w$ to rescale the intensity values, considering $c - \lfloor w/2 \rfloor  - \lfloor 0.25w \rfloor$ as the minimum value and $c + \lfloor w/2 \rfloor$ as the maximum value.

\textbf{PLANMED dataset.} In this dataset, we include 48 MLO view images acquired using the Planmed Nuance equipment. As in the IMS case, we also include an additional pre-processing operation due to the image format, which presents inverted values. Thus, we adopt the following minimum and maximum values to rescale the negative version of the input image: $x_{\text{min}} = -(c + \lfloor w/2 \rfloor + \lfloor 0.25w \rfloor )$ and $x_{\text{max}} = - (c - \lfloor w/2 \rfloor ) $. The rest of the pre-processing operations are the same as explained above. 

\textbf{HOLOGIC dataset.} This dataset includes a collection of 34 MLO view mammography images acquired using Selenia Dimensions equipment. In this case, these images follow the standard pre-processing pipeline, as in the GE case.

\subsection{Mammography image segmentation}
\label{sec:segmentation}

In this section, we present the reference approach proposed in \cite{sierra2024towards}, which modeled the problem as a semantic segmentation task that can be tackled using deep learning models. More precisely, we describe a baseline model and its corresponding training settings,  numerical results on the different datasets, and visual results useful to discuss the benefits and drawbacks of this method in the processing of mammography images of different vendors' equipment. 

\subsubsection{Model training}

While \cite{sierra2024towards} presents diverse experiments involving various deep learning model architectures and training configurations, this study adopts as a baseline a U-Net architecture in conjunction with an EfficientNetB3 model serving as a feature extractor (backbone). The network input consists of a single-channel image with dimensions $384\times384$, with intensity values in the range $[0,1]$. The network's output takes the form of a $384\times384\times C$ per-pixel probability map, where $C$ is the number of classes encompassing an implicit background class for unannotated pixels. Given that the segmentation task is treated as a multi-class per-pixel classification problem, the final layer incorporates a softmax activation function. For the training phase, we employ a hybrid loss function combining Categorical Focal Loss and Jaccard Loss functions, with a batch size of 4, a learning rate of $10^{-3}$, and a maximum of 200 epochs, integrating early stopping with patience of 30. The model is trained on the GE training set without considering augmentation operations and using the GE validation set to select the best weights regarding the loss function.

\subsubsection{Model evaluation}

To evaluate the model, we consider the datasets presented in the previous section that represent mammography images of different vendors' equipment. We use the metric Intersection Over Union (IoU), a widely used metric for semantic segmentation evaluation. This metric measures the degree of overlap between the segmentation prediction and the ground-truth segmentation (annotation). Thus, we can apply this metric to each class, obtaining IoU scores for each structure. 

Table \ref{tab:baseline_iou} shows the IoU results on the different datasets. As expected, the model presents good results on the test set of the GE dataset, similar to the results found in \cite{sierra2024towards}. The nipple seems to be the most challenging structure; however, it presents lower values because it is a small structure that tends to be more sensitive to the metrics. Thus, as expected, this is a good model for segmenting mammography images generated by GE equipment.

\begin{table}[!t]    \caption{Baseline approach IoU results}
    \tiny
    \begin{tabular}{ J{1.7cm} J{1cm} J{1cm} J{1cm} J{1cm}} 
    \hline
    \textbf{Dataset} & \textbf{Nipple} & \textbf{Pectoral} & \textbf{Fib. Tissue} & \textbf{Fat. Tissue}\\
    \hline
    GE & 0.7488 & 0.9608 & 0.9069 & 0.8078\\
    IMS & 0.7401 &	0.9165 & 0.7120 &	0.6070\\
    PLANMED & 0.7015 & 0.9432 & 0.7736 & 0.5962\\
    HOLOGIC & 0.1463 & 0.7677 & 0.6487 & 0.4192 \\
    \hline
    \end{tabular}
    \label{tab:baseline_iou}
\end{table}

In contrast, we can see that the results over the IMS, PLANMED, and HOLOGIC datasets are considerably worse on average, especially in the HOLOGIC case. These results suggest that the model lacks robust generalization across images by different vendor equipment. This is primarily due to the image differences which are not considered during training.

\begin{figure}[!t]
    \centering
    
    \includegraphics[width=\columnwidth]{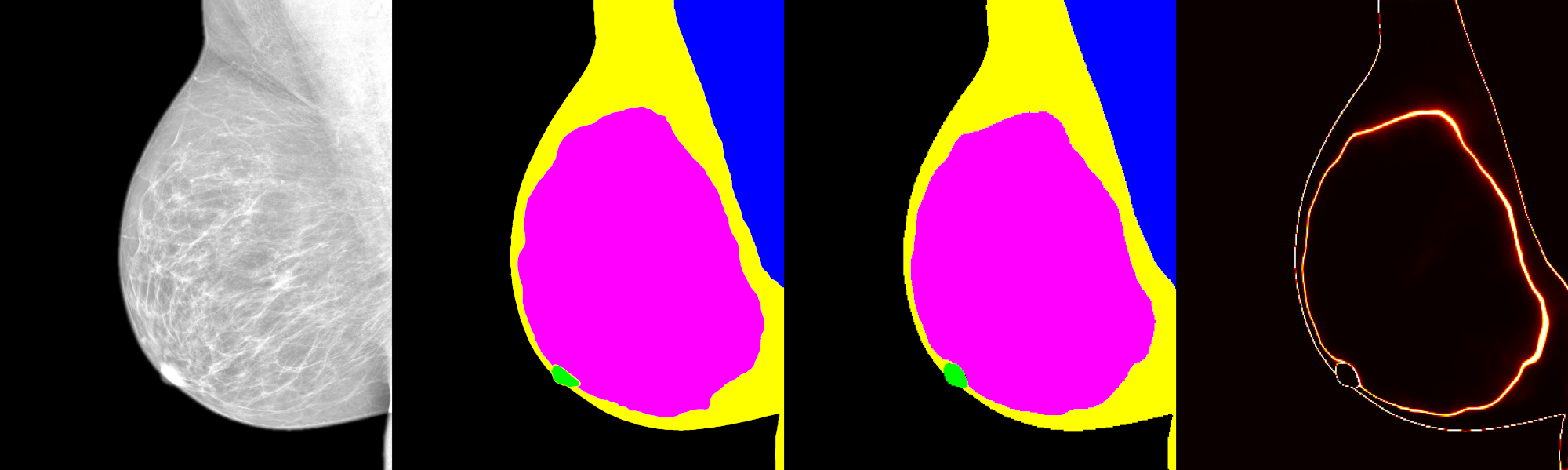}

    \includegraphics[width=\columnwidth]{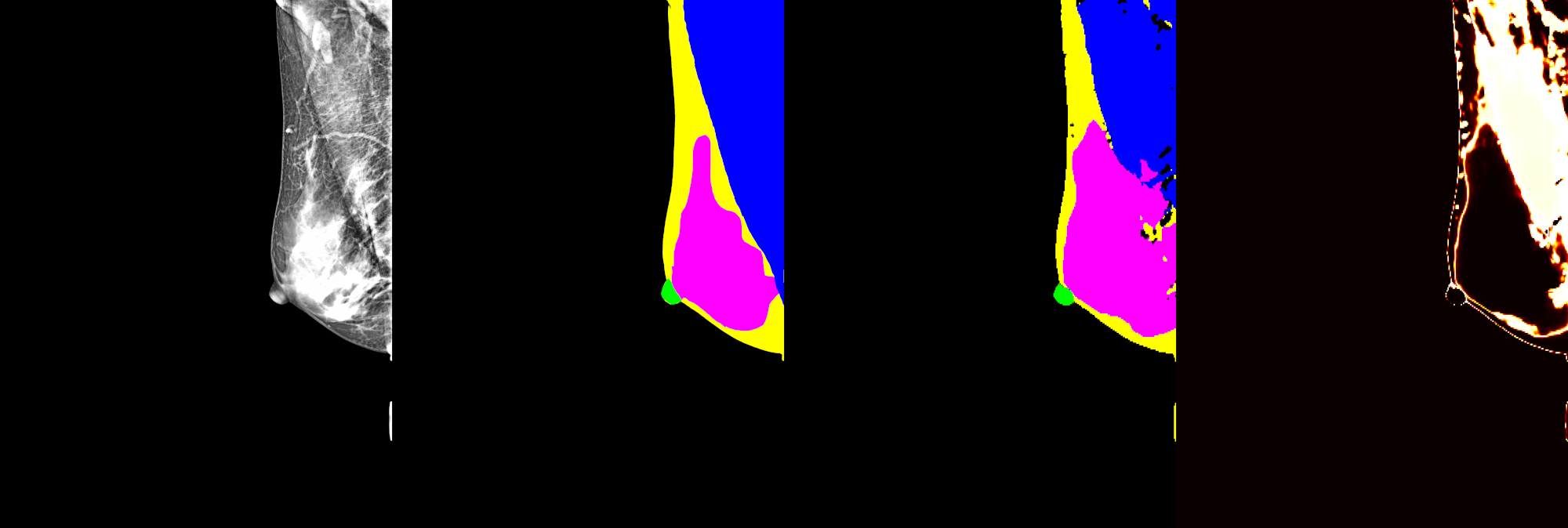}

    \includegraphics[width=\columnwidth]{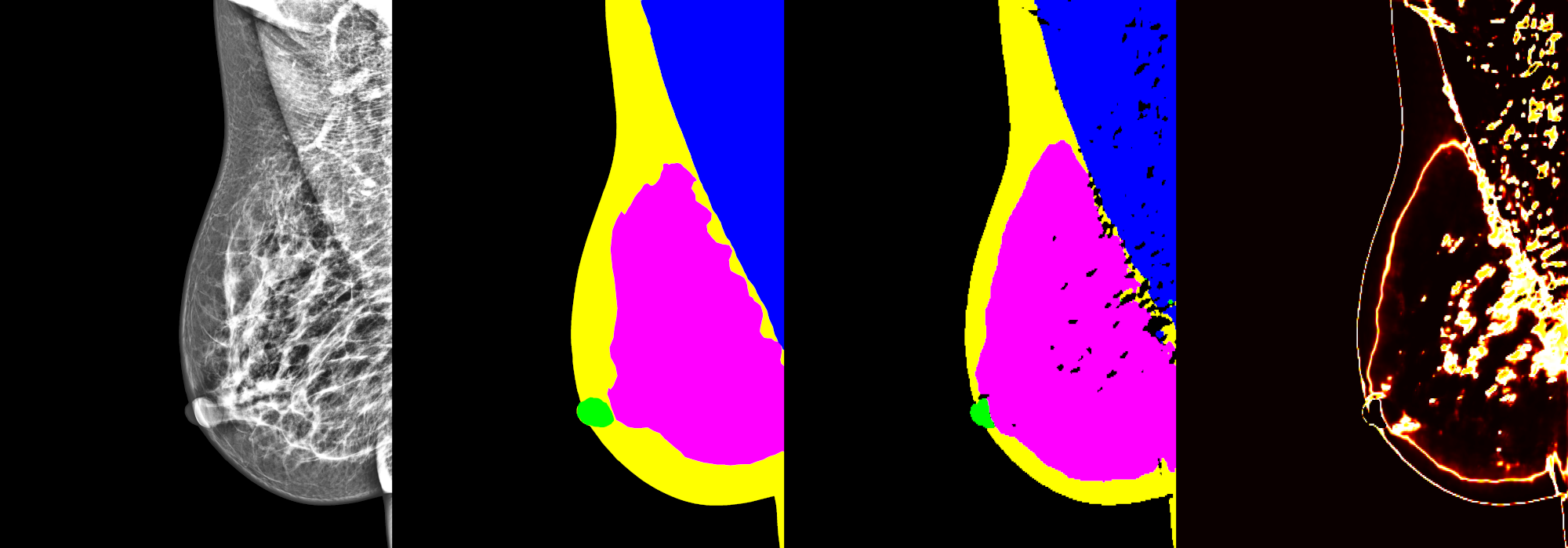}

    \includegraphics[width=\columnwidth]{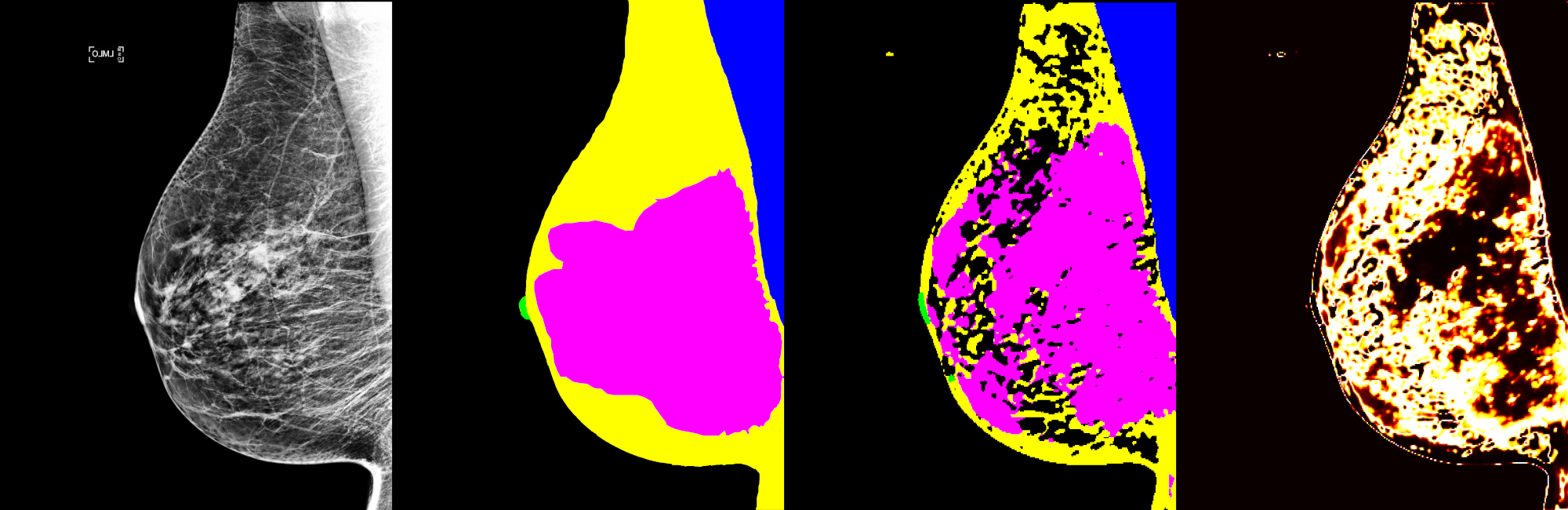}
    
    \caption{Baseline method visual results. First column: input image. Second column: ground-truth annotation. Third column: prediction. Fourth column: uncertainty map (hot color map with values in the range $[0,1]$). First row: image from GE dataset. Second row: image from IMS dataset. Third row: image from PLANMED dataset. Fourth row: image from HOLOGIC dataset.}
    \label{fig:baseline_initial_results}
\end{figure}

Figure \ref{fig:baseline_initial_results} shows some visual results on GE, IMS, PLANMED, and HOLOGIC images. Alongside the predicted structures, we include the uncertainty map generated using Test Time Augmentation (TTA). This map emphasizes areas of high model uncertainty, indicating the regions where the model predictions are least confident. 

In the GE case, we observe predictions closely aligned with ground-truth annotations, accompanied by well-defined uncertainty maps. As expected, the highlighted regions in the uncertainty maps closely match the prediction boundaries, indicating confident segmentation. Differently, we notice noisy predictions and chaotic uncertainty maps for other vendors, with thick highlighted regions in most cases. This noise in predictions and chaotic uncertainty map patterns suggest suboptimal model performance.


Using the model trained on the GE dataset reveals several limitations when dealing with images generated by other vendor equipment. The resulting predictions in these cases are often noisy with large high-uncertainty areas. Developing methods to enhance model generalization and adaptation across multiple vendor scenarios, avoiding the need for generating new manual annotations, could significantly contribute to its successful integration into clinical practice.
\subsection{Generalization improvement}
\label{sec:generalization}

In this section, we introduce two distinct methodologies for improving generalization, along with a combined approach. These methodologies involve augmenting training data samples to more accurately represent non-GE data by employing image intensity manipulation and style transfer techniques. A comprehensive explanation of these approaches follows.

\subsubsection{Image manipulation}

Data augmentation is usually related to applying random image transformations to the existing samples to achieve better generalization and robustness. The characteristics of the target domain guide the selection of these transformations we aim to represent. Thus, we propose a set of operations for manipulating image intensity values, enabling better representation of non-GE images. 

The general idea of our custom image augmentation procedure is to rescale the intensity values using the information of the annotated structures. Algorithm~\ref{alg:image_manipulation} summarizes this procedure, which receives as input an image $\mathbf{I}_\text{in}$, the binary mask $\mathbf{M}_\text{nip}$ of the nipple, the binary mask $\mathbf{M}_\text{fib}$ of the fibroglandular tissue, the binary mask $\mathbf{M}_\text{fat}$ of the fatty tissue, and the binary mask $\mathbf{M}_\text{b}$ of the background, and returns a manipulated version of $\mathbf{I}$, i.e. $\mathbf{I}_\text{out}$. 

\begin{algorithm}[!t]
\caption{Image manipulation}\label{alg:image_manipulation}
\scriptsize
\begin{algorithmic}[1]
\Procedure{manipulate}{$\mathbf{I}_\text{in}$,$\mathbf{M}_\text{nip}$,$\mathbf{M}_\text{fib}$,$\mathbf{M}_\text{fat}$,$\mathbf{M}_\text{b}$}
\State $\mathbf{I} = \text{rand}(0.8,1.2)*\mathbf{I}_\text{in}$
\If{$\text{rand}(0,1) < 0.5$}
\State \Return $\mathbf{I}$
\EndIf
\State $\mu_\text{nip} = \text{mean}(\mathbf{I},\mathbf{M}_\text{nip})$
\State $\mu_\text{fat} = \text{mean}(\mathbf{I},\mathbf{M}_\text{fat})$
\State $\mu_\text{fib} = \text{mean}(\mathbf{I},\mathbf{M}_\text{fib})$
\State $p_\text{fat} = \text{percentile}_5(\mathbf{I},\mathbf{M}_\text{fat})$
\State $a_\text{min} = \text{clip}(\text{rand}((p_\text{fat}-20),(p_\text{fat}+20)),0,255)$
\State $b = 0.7\mu_\text{fat} + 0.3\mu_\text{fib}$
\If{$a_\text{min} > (\mu_\text{nip}-5)$}
\State $a_\text{min} = \max(0,(\mu_\text{nip}-5))$
\ElsIf{$\text{rand}(0,1) < 0.5 \wedge (\mu_\text{nip}-5) < b$ }
\State $a_\text{min} = \text{rand}(\max(0,(\mu_\text{nip}-5)),\mu_\text{nip})$
\EndIf
\State $a_\text{max} = \text{percentile}_{98}(\mathbf{I})$
\State $\mathbf{I}_\text{out} = \text{rescale\_intensity}(\mathbf{I},(a_\text{min},a_\text{max}),(0,255))$
\If{$\text{rand}(0,1) < 0.5$}
\State $\mathbf{I}_\text{out}[\mathbf{M}_\text{b}] = 0$
\EndIf
\If{$\text{rand}(0,1) < 0.5$}
\State $\mathbf{I}_\text{out} = \text{add\_label}(\mathbf{I}_\text{out})$
\EndIf
\State \Return $\mathbf{I}_\text{out}$
\EndProcedure
\end{algorithmic}
\end{algorithm}

The function $\text{mean}(\mathbf{I},\mathbf{M})$ computes the mean intensity of $\mathbf{I}$ considering the values within the mask $\mathbf{M}$ only. The function $\text{percentile}_5(\mathbf{I},\mathbf{M})$ computes the 5th percentile of the values of $\mathbf{I}$ within the mask $\mathbf{M}$. The function $\text{percentile}_{98}(\mathbf{I})$ computes the 98th percentile of the intensity values of $\mathbf{I}$. The function $\text{rand}(x_\text{init},x_\text{end})$ generates a random float value between $x_\text{init}$ and $x_\text{end}$. The function $\text{clip}(x_\text{val},x_\text{init},x_\text{end})$ is the classic clip function that limits the value $x_\text{val}$ within the range $[x_\text{init},x_\text{end}]$. The function $\text{rescale\_intensity}(\mathbf{I},(x_\text{min},x_\text{max}),(y_\text{min},y_\text{max}))$ rescales the intensity values of $\mathbf{I}$ to the range $[y_\text{min},y_\text{max}]$, considering $x_\text{min}$ and $x_\text{max}$ the minimum and maximum values for $\mathbf{I}$, respectively. The operation $\mathbf{I}[\mathbf{M}] = x$ assigns the value $x$ to all elements of the image $\mathbf{I}$ that fall within the mask $\mathbf{M}$. Finally, the function $\text{add\_label}(\mathbf{I})$ adds a synthetic view label to the image, considering a random location close to the top-left corner. 

The intuition of this augmentation procedure is to achieve higher contrast within the breast region, simulating the behavior noticed in the non-GE equipment images. This manipulation uses local intensity statistics of the annotated structures to achieve robustness and avoid erasing regions of interest from the image, such as the nipple. Further, as shown in Figure~\ref{fig:baseline_initial_results}, HOLOGIC images always include a label describing laterality and view position. For this reason, we randomly add a synthetic label to simulate this case. Figure~\ref{fig:image_manipulation_example} shows some examples of our custom augmentation procedure.

\begin{figure}[!t]
    \centering
    \includegraphics[width=1.0\columnwidth]{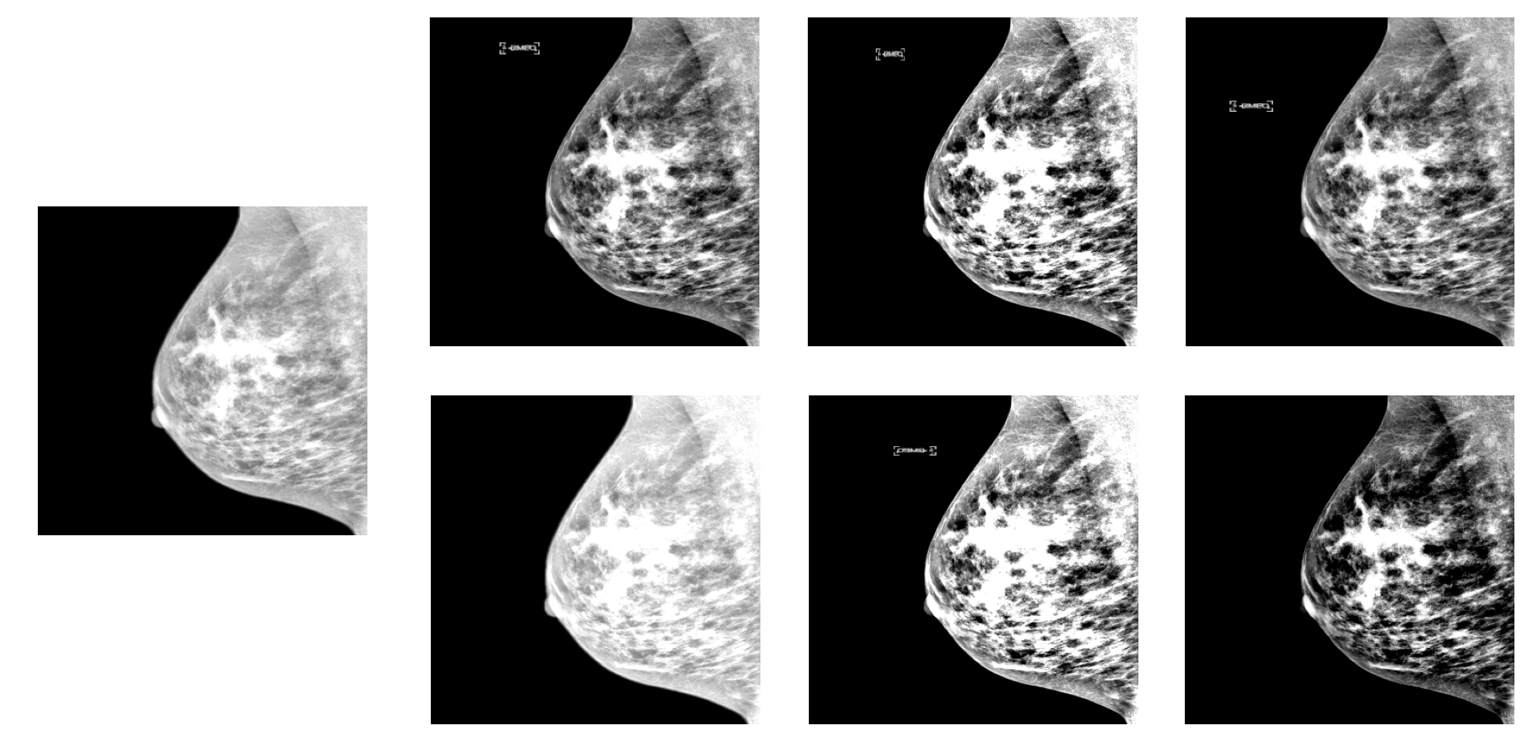}
    \caption{Image manipulation example. The most left image is the image $\mathbf{I}_\text{in}$. The other images are different results of applying the image manipulation algorithm.}
    \label{fig:image_manipulation_example}
\end{figure}

To train the segmentation model, we use the same settings described in Section~\ref{sec:segmentation} and include the custom image intensity manipulation procedure across all training and validation images. This annotation-guided augmentation method allows us to modify images in a context-aware manner, enhancing the model's ability to generalize across multiple vendor scenarios.


\subsubsection{Style transfer}

Style transfer synthesizes novel images by merging the content of one image with the style of another. Various deep learning frameworks provide pre-trained models for style transfer, which can be fine-tuned to specific styles. Once trained, these models can effectively transfer the learned style to any input image, serving as an effective tool for data augmentation. 

We aim to use style transfer to generate images resembling those from the non-GE equipment datasets, creating three different stylization models that adapt GE images to the IMS, PLANMED, and HOLOGIC styles. Then, using these models, we augment the training dataset to enhance generalization.

First, we select a reference image for each non-GE dataset, i.e. IMS, PLANMED, and HOLOGIC. Then, we fine-tune the model MLStyleTransfer from Apple's CreateML framework \cite{sahin2021introduction} to capture the style of each selected image. This fine-tuning process results in three distinct models, each capable of processing a $512\times512$ 3-channel image and producing a similarly dimensioned stylized output. The models are fine-tuned over 550 iterations, a style strength of 6, and a style density of 256. We validated the training process by visually assessing the stylized results on GE images.

\begin{figure}[!t]
    \centering
    \includegraphics[width=1.0\columnwidth]{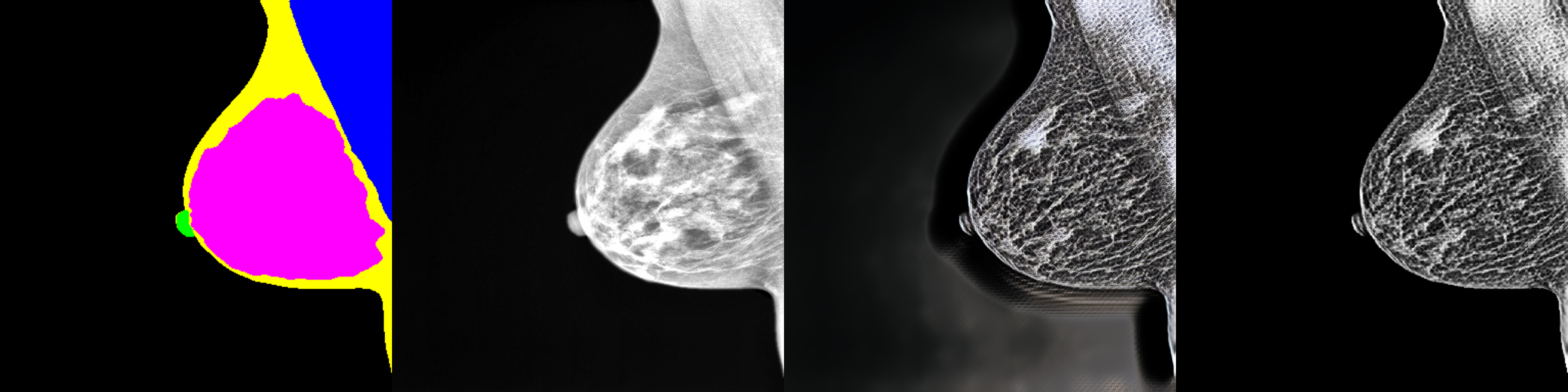}
    \caption{Style transfer post-processing. First column: annotated regions, where the background is colored in black. Second column: original image. Third column: stylized image. Fourth column: post-processed stylized image.}
    \label{fig:style_transfer_postprocessing}
\end{figure}

\begin{figure}[!t]
    \centering
    \includegraphics[width=1.0\columnwidth]{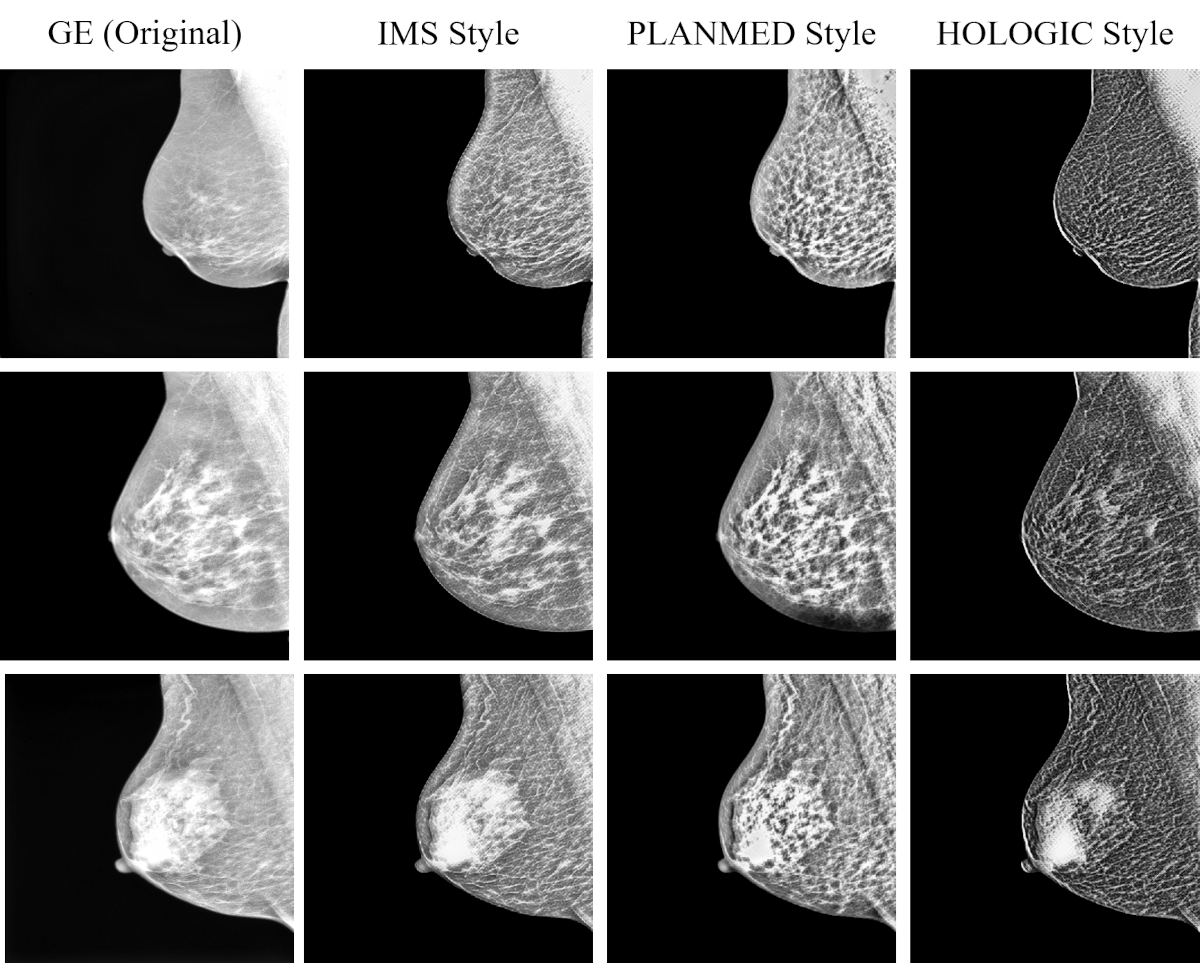}
    \caption{Style transfer examples. Each row is a different case. First column: original GE image. Second column: IMS stylization results. Third column: Second column: PLANMED stylization results. Second column: HOLOGIC stylization results.}
    \label{fig:style_transfer_examples}
\end{figure}

After training the models, we apply them to the entire GE training set to create synthetic images based on IMS, PLANMED, and HOLOGIC styles. To prepare these images for segmentation model training, we convert the stylized images into $384\times384$ single-channel images, which are the required input of our segmentation model. Additionally, to mitigate artifacts generated during the stylization process, we zero out all pixels within the annotated background region. Figure~\ref{fig:style_transfer_postprocessing} illustrates the post-processing operation, while Figure~\ref{fig:style_transfer_examples} shows examples of the stylization process using the three models.

To train the segmentation model using the stylized images, we keep the same settings described in Section~\ref{sec:segmentation}. Throughout the training process, we aim to achieve a balanced distribution between the original and various stylized image versions. As a result, each sampled input is equally likely to be either an original, IMS stylized image, PLANMED stylized image, or HOLOGIC stylized image, with a 25\% probability for each category. The same processing is applied to the validation set.

\subsubsection{Combining image manipulation with style transfer}

Both image manipulation and style transfer strategies offer unique advantages and drawbacks. Combining these methods can yield a more robust approach that enhances the segmentation model's generalization capabilities. We implement a straightforward combination by allocating a 20\% probability to each category of images: original, image manipulation results, and stylized images from IMS, PLANMED, and HOLOGIC datasets. As with previous cases, we preserve the same settings presented in Section~\ref{sec:segmentation} to train the segmentation model, applying this augmented approach to both the training and validation sets.
\section{Results}
\label{sec:results}

We use the datasets described in Section~\ref{sec:datasets} to compare the methods we proposed in this work with the baseline method outlined in Section~\ref{sec:segmentation}. We assess numerical performance using six standard metrics commonly used for evaluating semantic segmentation methods. These metrics include precision, recall, accuracy, Dice coefficient (F1-Score), IoU, and Hausdorff distance. Specifically, for the Hausdorff distance, we calculate the average of the two one-sided Hausdorff distances between the prediction and ground-truth structure contours in meters. We then present the metric values for each structure of interest by averaging these measurements across all tested images. We also present a mean value representing all the structures, excluding the background class. For visual analysis, we present the predictions and uncertainty maps as illustrated in Figure~\ref{fig:baseline_initial_results}. 

Tables \ref{tab:numerical_results_ge}, \ref{tab:numerical_results_ims}, \ref{tab:numerical_results_planmed}, and \ref{tab:numerical_results_hologic} present the numerical results for the GE, IMS, PLANMED, and HOLOGIC datasets, while Figures~\ref{fig:ge_results_vis}, ~\ref{fig:ims_results_vis}, ~\ref{fig:planmed_results_vis}, and ~\ref{fig:hologic_results_vis} illustrates the corresponding visual results. Additionally, Tables \ref{tab:numerical_results_ge_ss}, \ref{tab:numerical_results_ims_ss}, \ref{tab:numerical_results_planmed_ss}, and \ref{tab:numerical_results_hologic_ss} present the pairs of methods that exhibit statistically significant differences for each dataset, metric, and structure of interest. Comparisons not included in these tables do not show significant differences. This analysis was conducted using the Kruskal-Wallis test, followed by the Dunn method with Bonferroni adjustment to identify specific method pairs with significant differences, considering $p < 0.05$.

The numerical results from the GE dataset (Table~\ref{tab:numerical_results_ge}) show that the proposed data augmentation methods perform comparably to the baseline method. This results suggests that the proposed approaches do not degrade performance on this type of images. Moreover, they yield improved segmentation for the pectoral muscle, which is one of the most important structures for mammography positioning analysis. Figure~\ref{fig:ge_results_vis} illustrates the consistent quality of predictions across various anatomies, even in complex cases where the nipple overlap other tissues.

\begin{table*}[t]    \caption{Numerical results on the GE dataset (test)}
    \tiny
    \begin{center}
    \begin{tabular}{ J{2cm} J{3cm} J{1.2cm} J{1.2cm} J{1.2cm} J{1.2cm} J{1.2cm}} 
    \hline
    \textbf{Metric} & \textbf{Method} & \textbf{Nipple} & \textbf{Pectoral} & \textbf{Fib. Tissue} & \textbf{Fat. Tissue} & \textbf{Mean}\\
    \hline
    \multirow{4}{*}{Precision} & Baseline & 0.8349 & \textbf{0.9909} & 0.9491 & \textbf{0.8990} & \textbf{0.9185} \\
    & Image manipulation & \textbf{0.8441} & 0.9830 & 0.9557 & 0.8892 & 0.9180 \\
    & Style transfer & 0.8212 & 0.9866 & \textbf{0.9614} & 0.8707 & 0.9100 \\
    & Combination & 0.8409 & 0.9871 & 0.9443 & 0.8972 & 0.9174 \\
    \hline
    \multirow{4}{*}{Recall} & Baseline & \textbf{0.8867} & 0.9695 & 0.9543 & 0.8880 & 0.9246 \\
    & Image manipulation & 0.8632 & \textbf{0.9799} & 0.9479 & 0.9068 & 0.9244 \\
    & Style transfer & 0.8807 & 0.9773 & 0.9341 & \textbf{0.9165} & \textbf{0.9272} \\
    & Combination & 0.8610 & 0.9741 & \textbf{0.9546} & 0.8890 & 0.9197 \\
    \hline
    \multirow{4}{*}{Accuracy} & Baseline & \textbf{0.9995} & 0.9971 & 0.9798 & 0.9757 & 0.9880 \\
    & Image manipulation & 0.9994 & 0.9972 & \textbf{0.9799} & \textbf{0.9760} & \textbf{0.9881} \\
    & Style transfer & 0.9994 & 0.9972 & 0.9783 & 0.9743 & 0.9873 \\
    & Combination & 0.9994 & \textbf{0.9972} & 0.9786 & 0.9750 & 0.9876 \\
    \hline
    \multirow{4}{*}{Dice} & Baseline & \textbf{0.8464} & 0.9780 & 0.9496 & 0.8882 & \textbf{0.9156} \\
    & Image manipulation & 0.8367 & 0.9799 & \textbf{0.9497} & \textbf{0.8931} & 0.9149 \\
    & Style transfer & 0.8344 & \textbf{0.9808} & 0.9450 & 0.8877 & 0.9120 \\
    & Combination & 0.8358 & 0.9789 & 0.9473 & 0.8878 & 0.9124 \\
    \hline
    \multirow{4}{*}{IoU} & Baseline & \textbf{0.7488} & 0.9608 & 0.9069 & 0.8078 & \textbf{0.8561} \\
    & Image manipulation & 0.7344 & 0.9634 & \textbf{0.9070} & \textbf{0.8150} & 0.8550 \\
    & Style transfer & 0.7316 & \textbf{0.9644} & 0.8988 & 0.8061 & 0.8502 \\
    & Combination & 0.7333 & 0.9623 & 0.9024 & 0.8061 & 0.8510 \\
    \hline
    \multirow{4}{*}{Hausdorff} & Baseline & \textbf{0.0019} & 0.0038 & 0.0192 & 0.0141 & 0.0098 \\
    & Image manipulation & 0.0020 & 0.0046 & \textbf{0.0106} & \textbf{0.0134} & \textbf{0.0076} \\
    & Style transfer & 0.0020 & \textbf{0.0036} & 0.0110 & 0.0137 & 0.0076 \\
    & Combination & 0.0020 & 0.0039 & 0.0110 & 0.0139 & 0.0077 \\
    \hline
    \end{tabular}
    \end{center}
    \label{tab:numerical_results_ge}
\end{table*}

\begin{table*}[t]    \caption{Statistical significance of the difference in results on the GE dataset (test)}
    \tiny
    \begin{center}
    \begin{tabular}{ J{1cm} J{1cm} J{13cm}} 
    \hline
    \textbf{Metric} & \textbf{Structure} & \textbf{Pairs} \\
    \hline
     Precision & Nipple & (Image manipulation, Style transfer), (Style transfer, Combination) \\ \hline
     Precision & Pectoral & (Baseline, Image manipulation), (Baseline, Style transfer), (Baseline, Combination), (Image manipulation, Combination), (Style transfer, Combination) \\ \hline
     Precision & Fib. Tissue & (Baseline, Image manipulation), (Baseline, Style transfer), (Image manipulation, Style transfer), (Image manipulation, Combination), (Style transfer, Combination) \\ \hline
     Precision & Fat. Tissue & (Baseline, Style transfer), (Image manipulation, Style transfer), (Style transfer, Combination) \\ \hline
     Precision & Mean & (Baseline, Style transfer), (Image manipulation, Style transfer), (Style transfer, Combination) \\ \hline
     Recall & Nipple & (Baseline, Image manipulation), (Baseline, Combination) \\ \hline
     Recall & Pectoral & (Baseline, Image manipulation), (Baseline, Style transfer), (Baseline, Combination), (Image manipulation, Style transfer), (Image manipulation, Combination) \\ \hline
     Recall & Fib. Tissue & (Baseline, Image manipulation), (Baseline, Style transfer), (Image manipulation, Style transfer), (Image manipulation, Combination), (Style transfer, Combination) \\ \hline
     Recall & Fat. Tissue & (Baseline, Image manipulation), (Baseline, Style transfer), (Image manipulation, Style transfer), (Image manipulation, Combination), (Style transfer, Combination) \\ \hline
     Accuracy & Pectoral & (Baseline, Image manipulation), (Baseline, Combination) \\ \hline
     Dice & Pectoral & (Baseline, Image manipulation), (Baseline, Combination) \\ \hline
     Dice & Fib. Tissue & (Baseline, Style transfer), (Image manipulation, Style transfer) \\ \hline
     IoU & Pectoral & (Baseline, Image manipulation), (Baseline, Combination) \\ \hline
     IoU & Fib. Tissue & (Baseline, Style transfer), (Image manipulation, Style transfer) \\ \hline
     Hausdorff & Fib. Tissue & (Baseline, Image manipulation), (Baseline, Style transfer), (Baseline, Combination) \\ \hline
    \end{tabular}
    \end{center}
    \label{tab:numerical_results_ge_ss}
\end{table*}

\begin{figure*}[t]
    \centering
    \includegraphics[width=0.65\textwidth]{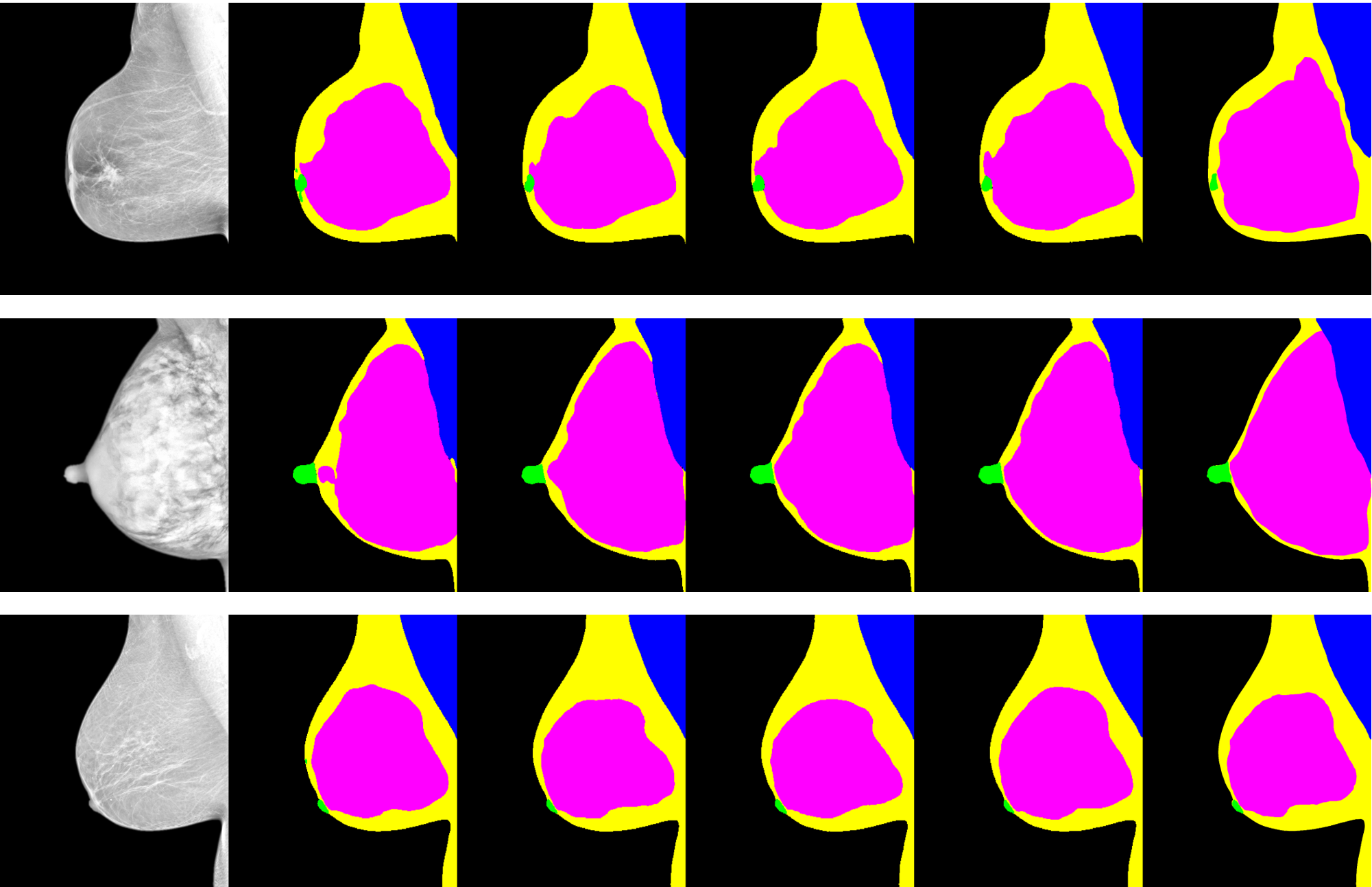}
    \caption{Visual results on the GE dataset (test). Each row represents a different case. First column: input image. Second column: baseline result. Third column: image manipulation result. Fourth column: style transfer result. Fifth column: image manipulation and style transfer combination result. Sixth column: ground-truth annotation.}
    \label{fig:ge_results_vis}
\end{figure*}

The image manipulation method presents superior mean numerical results than the baseline on the PLANMED and IMS datasets, as shown in Tables~\ref{tab:numerical_results_ims} and~\ref{tab:numerical_results_planmed}. However, there is a noticeable decline in performance for the nipple segmentation. Thus, this method may not be optimal for applications necessitating accurate detection and delineation of the nipple in these images. In the case of the HOLOGIC dataset, which is the most distant from the GE dataset in terms of image similarity, we can notice superior performance when compared to both the baseline and style transfer methods. This is more evident in the nipple structure, where the Dice, IoU, and Hausdorff metrics highlight this superiority. Further, as shown in Figure~\ref{fig:hologic_results_vis}, the synthetic labels added during training are helpful in classifying the HOLOGIC image labels as background instead of breast tissues.

\begin{table*}[t]    \caption{Numerical results on the IMS dataset}
    \tiny
    \begin{center}
    \begin{tabular}{ J{2cm} J{3cm} J{1.2cm} J{1.2cm} J{1.2cm} J{1.2cm} J{1.2cm}} 
    \hline
    \textbf{Metric} & \textbf{Method} & \textbf{Nipple} & \textbf{Pectoral} & \textbf{Fib. Tissue} & \textbf{Fat. Tissue} & \textbf{Mean}\\
    \hline
    \multirow{4}{*}{Precision} & Baseline & \textbf{0.9139} & \textbf{0.9971} & 0.7207 & 0.9149 & 0.8867 \\
    & Image manipulation & 0.8220 & 0.9853 & 0.8021 & 0.8941 & 0.8759 \\
    & Style transfer & 0.8859 & 0.9779 & \textbf{0.8768} & 0.9159 & \textbf{0.9142} \\
    & Combination & 0.8981 & 0.9853 & 0.8226 & \textbf{0.9294} & 0.9088 \\
    \hline
    \multirow{4}{*}{Recall} & Baseline & 0.7979 & 0.9191 & \textbf{0.9855} & 0.6411 & 0.8359 \\
    & Image manipulation & 0.7871 & 0.9771 & 0.9837 & 0.7805 & 0.8821 \\
    & Style transfer & \textbf{0.8844} & \textbf{0.9824} & 0.9537 & \textbf{0.8646} & \textbf{0.9213} \\
    & Combination & 0.8290 & 0.9765 & 0.9815 & 0.8043 & 0.8978 \\
    \hline
    \multirow{4}{*}{Accuracy} & Baseline & 0.9995 & 0.9943 & 0.9692 & 0.9638 & 0.9817 \\
    & Image manipulation & 0.9994 & 0.9975 & 0.9802 & 0.9738 & 0.9877 \\
    & Style transfer & 0.9996 & 0.9974 & \textbf{0.9860} & \textbf{0.9818} & \textbf{0.9912} \\
    & Combination & \textbf{0.9996} & \textbf{0.9975} & 0.9824 & 0.9781 & 0.9894 \\
    \hline
    \multirow{4}{*}{Dice} & Baseline & 0.8473 & 0.9540 & 0.8287 & 0.7506 & 0.8451 \\
    & Image manipulation & 0.7925 & \textbf{0.9806} & 0.8807 & 0.8313 & 0.8713 \\
    & Style transfer & \textbf{0.8794} & 0.9797 & \textbf{0.9103} & \textbf{0.8876} & \textbf{0.9143} \\
    & Combination & 0.8545 & 0.9803 & 0.8916 & 0.8598 & 0.8965 \\
    \hline
    \multirow{4}{*}{IoU} & Baseline & 0.7401 &	0.9165 & 0.7120 &	0.6070 & 0.7439 \\
    & Image manipulation & 0.6679 & \textbf{0.9628} & 0.7902 & 0.7150 & 0.7840 \\
    & Style transfer & \textbf{0.7904} & 0.9608 & \textbf{0.8378} & \textbf{0.8001} & \textbf{0.8473} \\
    & Combination & 0.7545 & 0.9621 & 0.8080 & 0.7573 & 0.8205 \\
    \hline
    \multirow{4}{*}{Hausdorff} & Baseline & 0.0123 & 0.0148 & 0.0346 & 0.0253 & 0.0218 \\
    & Image manipulation & \textbf{0.0025} & \textbf{0.0068} & 0.0156 & 0.0210 & \textbf{0.0115} \\
    & Style transfer & 0.0227 & 0.0095 & \textbf{0.0125} & \textbf{0.0190} & 0.0159 \\
    & Combination & 0.0030 & 0.0078 & 0.0157 & 0.0209 & 0.0118 \\
    \hline
    \end{tabular}
    \end{center}
    \label{tab:numerical_results_ims}
\end{table*}

\begin{table*}[t]    \caption{Statistical significance of the difference in results on the IMS dataset}
    \tiny
    \begin{center}
    \begin{tabular}{ J{1cm} J{1cm} J{13cm}} 
    \hline
    \textbf{Metric} & \textbf{Structure} & \textbf{Pairs} \\
    \hline
    Precision & Nipple & (Baseline, Image manipulation), (Image manipulation, Style transfer), (Image manipulation, Combination) \\ \hline
    Precision & Pectoral & (Baseline, Image manipulation), (Baseline, Style transfer), (Baseline, Combination) \\ \hline
    Precision & Fib. Tissue & (Baseline, Image manipulation), (Baseline, Style transfer), (Baseline, Combination), (Image manipulation, Style transfer), (Style transfer, Combination) \\ \hline
    Precision & Fat. Tissue & (Baseline, Image manipulation), (Image manipulation, Combination) \\ \hline
    Precision & Mean & (Baseline, Style transfer), (Baseline, Combination), (Image manipulation, Style transfer), (Image manipulation, Combination) \\ \hline
    Recall & Nipple & (Baseline, Style transfer), (Image manipulation, Style transfer) \\ \hline
    Recall & Pectoral & (Baseline, Image manipulation), (Baseline, Style transfer), (Baseline, Combination) \\ \hline
    Recall & Fib. Tissue & (Baseline, Style transfer), (Image manipulation, Style transfer), (Style transfer, Combination) \\ \hline
    Recall & Fat. Tissue & (Baseline, Image manipulation), (Baseline, Style transfer), (Baseline, Combination), (Image manipulation, Style transfer), (Style transfer, Combination) \\ \hline
    Recall & Mean & (Baseline, Image manipulation), (Baseline, Style transfer), (Baseline, Combination), (Image manipulation, Style transfer), (Style transfer, Combination) \\ \hline
    Accuracy & Nipple & (Baseline, Image manipulation), (Image manipulation, Style transfer), (Image manipulation, Combination) \\ \hline
    Accuracy & Fib. Tissue & (Baseline, Image manipulation), (Baseline, Style transfer), (Baseline, Combination), (Image manipulation, Style transfer) \\ \hline
    Accuracy & Fat. Tissue & (Baseline, Image manipulation), (Baseline, Style transfer), (Baseline, Combination), (Image manipulation, Style transfer) \\ \hline
    Accuracy & Mean & (Baseline, Image manipulation), (Baseline, Style transfer), (Baseline, Combination), (Image manipulation, Style transfer) \\ \hline
    Dice & Nipple & (Baseline, Image manipulation), (Baseline, Style transfer), (Image manipulation, Style transfer), (Image manipulation, Combination) \\ \hline
    Dice & Pectoral & (Baseline, Image manipulation), (Baseline, Combination) \\ \hline
    Dice & Fib. Tissue & (Baseline, Image manipulation), (Baseline, Style transfer), (Baseline, Combination), (Image manipulation, Style transfer) \\ \hline
    Dice & Fat. Tissue & (Baseline, Image manipulation), (Baseline, Style transfer), (Baseline, Combination), (Image manipulation, Style transfer) \\ \hline
    Dice & Mean & (Baseline, Image manipulation), (Baseline, Style transfer), (Baseline, Combination), (Image manipulation, Style transfer), (Image manipulation, Combination), (Style transfer, Combination) \\ \hline
    IoU & Nipple & (Baseline, Image manipulation), (Baseline, Style transfer), (Image manipulation, Style transfer), (Image manipulation, Combination) \\ \hline
    IoU & Pectoral & (Baseline, Image manipulation), (Baseline, Combination) \\ \hline
    IoU & Fib. Tissue & (Baseline, Image manipulation), (Baseline, Style transfer), (Baseline, Combination), (Image manipulation, Style transfer) \\ \hline
    IoU & Fat. Tissue & (Baseline, Image manipulation), (Baseline, Style transfer), (Baseline, Combination), (Image manipulation, Style transfer) \\ \hline
    IoU & Mean & (Baseline, Image manipulation), (Baseline, Style transfer), (Baseline, Combination), (Image manipulation, Style transfer), (Image manipulation, Combination) \\ \hline
    Hausdorff & Nipple & (Baseline, Combination) \\ \hline
    Hausdorff & Pectoral & (Baseline, Image manipulation), (Baseline, Combination) \\ \hline
    Hausdorff & Fib. Tissue & (Baseline, Image manipulation), (Baseline, Style transfer), (Baseline, Combination) \\ \hline
    Hausdorff & Fat. Tissue & (Baseline, Style transfer) \\ \hline
    Hausdorff & Mean & (Baseline, Image manipulation), (Baseline, Style transfer), (Baseline, Combination) \\ \hline
    \end{tabular}
    \end{center}
    \label{tab:numerical_results_ims_ss}
\end{table*}

\begin{figure*}[t]
    \centering
    \includegraphics[width=0.65\textwidth]{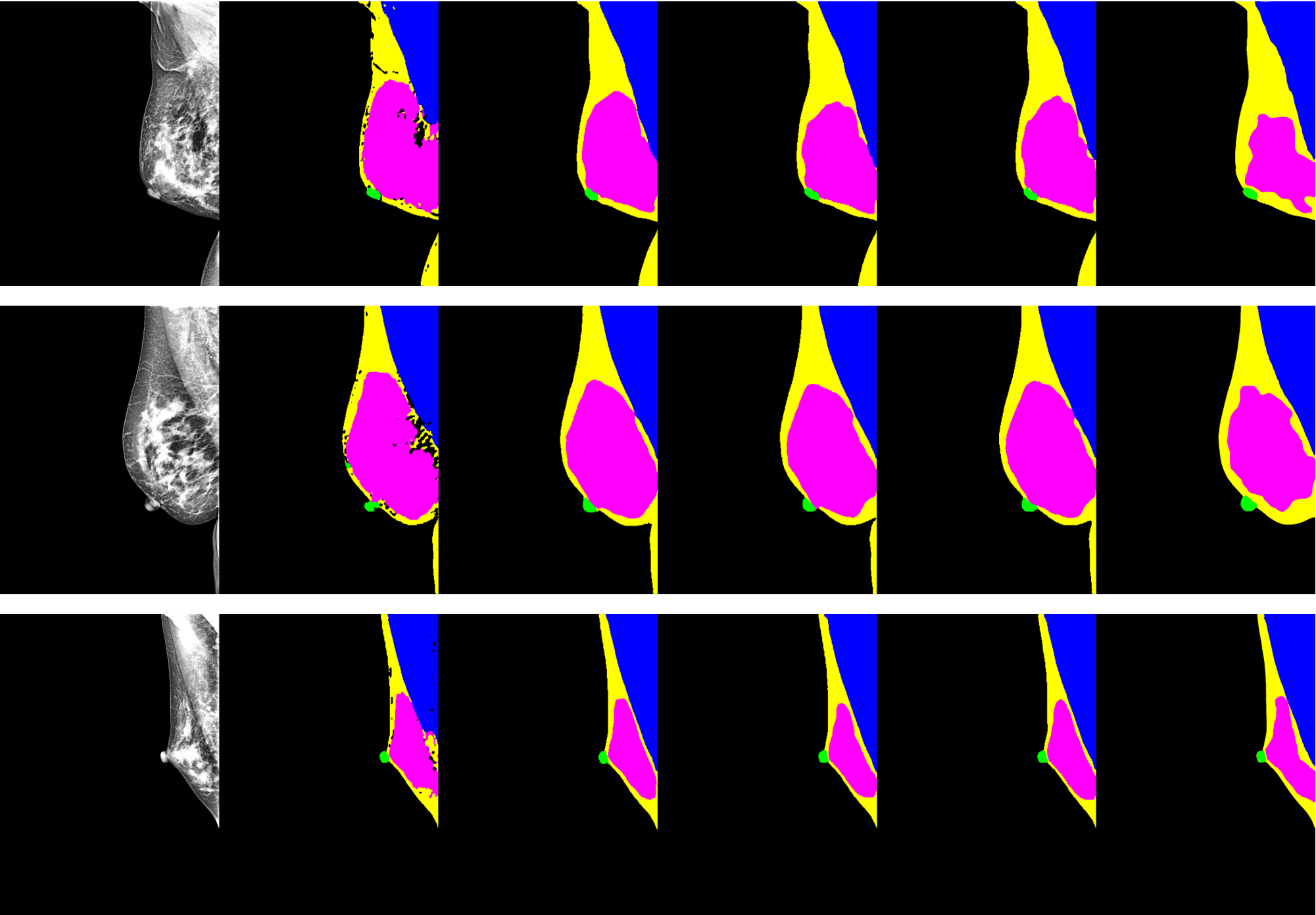}
    \caption{Visual results on the IMS dataset. Each row represents a different case. First column: input image. Second column: baseline result. Third column: image manipulation result. Fourth column: style transfer result. Fifth column: image manipulation and style transfer combination result. Sixth column: ground-truth annotation.}
    \label{fig:ims_results_vis}
\end{figure*}

The style transfer method presents superior performance on the IMS and PLANMED datasets, achieving the best IoU and Dice values for the nipple, fibroglandular tissue, and fatty tissue. For the pectoral muscle, the method presents results similar to its combination with the image manipulation method. In the case of the nipple, the Hausdorff distance is not the minimum because this method tends to create noisy nipple regions in challenging cases, such as the example shown in the last row of Figure~\ref{fig:planmed_results_vis}. While these noisy regions impact distance-based metrics like the Hausdorff distance, they do not represent an extensive area and can be easily removed in post-processing. Even with this superiority on IMS and PLANNED datasets, the style transfer method presents inferior results compared to the image manipulation method on the HOLOGIC dataset. In this case, we can see significantly lower IoU values and higher Hausdorff distances for the nipple structure, potentially affecting applications where accurate nipple localization is critical. Figure~\ref{fig:hologic_results_vis} corroborates the latter, where an image label region is misclassified as nipple and fatty tissue instead of background.

\begin{table*}[t]    \caption{Numerical results on the PLANMED dataset}
    \tiny
    \begin{center}
    \begin{tabular}{ J{2cm} J{3cm} J{1.2cm} J{1.2cm} J{1.2cm} J{1.2cm} J{1.2cm}} 
    \hline
    \textbf{Metric} & \textbf{Method} & \textbf{Nipple} & \textbf{Pectoral} & \textbf{Fib. Tissue} & \textbf{Fat. Tissue} & \textbf{Mean}\\
    \hline
    \multirow{4}{*}{Precision} & Baseline & \textbf{0.9463} & \textbf{0.9826} & 0.7816 & \textbf{0.9688} & 0.9198 \\
    & Image manipulation & 0.7940 & 0.9691 & 0.8525 & 0.9143 & 0.8825 \\
    & Style transfer & 0.9204 & 0.9705 & \textbf{0.9001} & 0.9116 & 0.9256 \\
    & Combination & 0.9291 & 0.9790 & 0.8661 & 0.9329 & \textbf{0.9268} \\
    \hline
    \multirow{4}{*}{Recall} & Baseline & 0.7352 & 0.9597 & \textbf{0.9878} & 0.6091 & 0.8229 \\
    & Image manipulation & 0.5990 & 0.9901 & 0.9715 & 0.7863 & 0.8367 \\
    & Style transfer & \textbf{0.8119} & \textbf{0.9903} & 0.9357 & \textbf{0.8495} & \textbf{0.8969} \\
    & Combination & 0.7834 & 0.9866 & 0.9656 & 0.8027 & 0.8846 \\
    \hline
    \multirow{4}{*}{Accuracy} & Baseline & 0.9989 & 0.9946 & 0.9493 & 0.9422 & 0.9713 \\
    & Image manipulation & 0.9982 & 0.9959 & 0.9668 & 0.9607 & 0.9804 \\
    & Style transfer & \textbf{0.9991} & 0.9963 & \textbf{0.9713} & \textbf{0.9674} & \textbf{0.9835} \\
    & Combination & 0.9990 & \textbf{0.9965} & 0.9689 & 0.9648 & 0.9823 \\
    \hline
    \multirow{4}{*}{Dice} & Baseline & 0.8132 & 0.9700 & 0.8693 & 0.7395 & 0.8480 \\
    & Image manipulation & 0.6549 & 0.9789 & 0.9061 & 0.8420 & 0.8455 \\
    & Style transfer & \textbf{0.8506} & 0.9798 & \textbf{0.9151} & \textbf{0.8753} & \textbf{0.9052} \\
    & Combination & 0.8327 & \textbf{0.9824} & 0.9113 & 0.8592 & 0.8964 \\
    \hline
    \multirow{4}{*}{IoU} & Baseline & 0.7015 & 0.9432 & 0.7736 & 0.5962 & 0.7536 \\
    & Image manipulation & 0.5198 & 0.9596 & 0.8308 & 0.7306 & 0.7602 \\
    & Style transfer & \textbf{0.7521} & 0.9611 & \textbf{0.8455} & \textbf{0.7809} & \textbf{0.8349} \\
    & Combination & 0.7325 & \textbf{0.9659} & 0.8389 & 0.7560 & 0.8233 \\
    \hline
    \multirow{4}{*}{Hausdorff} & Baseline & 0.0196 & 0.0170 & 0.0292 & 0.0214 & 0.0218 \\
    & Image manipulation & 0.0059 & 0.0089 & 0.0169 & 0.0184 & 0.0126 \\
    & Style transfer & 0.0073 & 0.0075 & \textbf{0.0147} & \textbf{0.0172} & 0.0117 \\
    & Combination & \textbf{0.0025} & \textbf{0.0071} & 0.0167 & 0.0185 & \textbf{0.0112} \\
    \hline
    \end{tabular}
    \end{center}
    \label{tab:numerical_results_planmed}
\end{table*}

\begin{table*}[t]    \caption{Statistical significance of the difference in results on the PLANMED dataset}
    \tiny
    \begin{center}
    \begin{tabular}{ J{1cm} J{1cm} J{13cm}} 
    \hline
    \textbf{Metric} & \textbf{Structure} & \textbf{Pairs} \\
    \hline
    Precision & Nipple & (Baseline, Image manipulation), (Image manipulation, Style transfer), (Image manipulation, Combination) \\ \hline
    Precision & Pectoral & (Baseline, Image manipulation), (Baseline, Style transfer), (Style transfer, Combination) \\ \hline
    Precision & Fib. Tissue & (Baseline, Image manipulation), (Baseline, Style transfer), (Baseline, Combination), (Image manipulation, Style transfer) \\ \hline
    Precision & Fat. Tissue & (Baseline, Image manipulation), (Baseline, Style transfer), (Baseline, Combination) \\ \hline
    Precision & Mean & (Baseline, Image manipulation), (Image manipulation, Style transfer), (Image manipulation, Combination) \\ \hline
    Recall & Nipple & (Image manipulation, Style transfer), (Image manipulation, Combination) \\ \hline
    Recall & Pectoral & (Baseline, Image manipulation), (Baseline, Style transfer), (Baseline, Combination) \\ \hline
    Recall & Fib. Tissue & (Baseline, Image manipulation), (Baseline, Style transfer), (Baseline, Combination), (Image manipulation, Style transfer), (Style transfer, Combination) \\ \hline
    Recall & Fat. Tissue & (Baseline, Image manipulation), (Baseline, Style transfer), (Baseline, Combination), (Image manipulation, Style transfer) \\ \hline
    Recall & Mean & (Baseline, Style transfer), (Baseline, Combination), (Image manipulation, Style transfer), (Image manipulation, Combination) \\ \hline
    Accuracy & Nipple & (Baseline, Image manipulation), (Image manipulation, Style transfer), (Image manipulation, Combination) \\ \hline
    Accuracy & Pectoral & (Baseline, Combination) \\ \hline
    Accuracy & Fib. Tissue & (Baseline, Image manipulation), (Baseline, Style transfer), (Baseline, Combination) \\ \hline
    Accuracy & Fat. Tissue & (Baseline, Image manipulation), (Baseline, Style transfer), (Baseline, Combination) \\ \hline
    Accuracy & Mean & (Baseline, Style transfer), (Baseline, Combination) \\ \hline
    Dice & Nipple & (Baseline, Image manipulation), (Image manipulation, Style transfer), (Image manipulation, Combination) \\ \hline
    Dice & Pectoral & (Baseline, Combination) \\ \hline
    Dice & Fib. Tissue & (Baseline, Image manipulation), (Baseline, Style transfer), (Baseline, Combination) \\ \hline
    Dice & Fat. Tissue & (Baseline, Image manipulation), (Baseline, Style transfer), (Baseline, Combination), (Image manipulation, Style transfer) \\ \hline
    Dice & Mean & (Baseline, Style transfer), (Baseline, Combination), (Image manipulation, Style transfer), (Image manipulation, Combination) \\ \hline
    IoU & Nipple & (Baseline, Image manipulation), (Image manipulation, Style transfer), (Image manipulation, Combination) \\ \hline
    IoU & Pectoral & (Baseline, Combination) \\ \hline
    IoU & Fib. Tissue & (Baseline, Image manipulation), (Baseline, Style transfer), (Baseline, Combination) \\ \hline
    IoU & Fat. Tissue & (Baseline, Image manipulation), (Baseline, Style transfer), (Baseline, Combination), (Image manipulation, Style transfer) \\ \hline
    IoU & Mean & (Baseline, Style transfer), (Baseline, Combination), (Image manipulation, Style transfer), (Image manipulation, Combination) \\ \hline
    Hausdorff & Nipple & (Baseline, Combination), (Image manipulation, Style transfer), (Image manipulation, Combination) \\ \hline
    Hausdorff & Pectoral & (Baseline, Image manipulation), (Baseline, Style transfer), (Baseline, Combination) \\ \hline
    Hausdorff & Fib. Tissue & (Baseline, Image manipulation), (Baseline, Style transfer), (Baseline, Combination) \\ \hline
    Hausdorff & Fat. Tissue & (Baseline, Style transfer), (Baseline, Combination) \\ \hline
    Hausdorff & Mean & (Baseline, Image manipulation), (Baseline, Style transfer), (Baseline, Combination) \\ \hline
    \end{tabular}
    \end{center}
    \label{tab:numerical_results_planmed_ss}
\end{table*}

The combination method offers a balance between the two approaches. When considering the full metrics across all four test datasets, we observe that this method consistently achieves the best or near-best numerical results. It leverages the strong generalization capabilities of the image manipulation method on HOLOGIC images while benefiting from the effective generalization of the style transfer method on IMS and PLANMED images. Further, from Figures~\ref{fig:ge_results_vis}, ~\ref{fig:ims_results_vis}, ~\ref{fig:planmed_results_vis}, and ~\ref{fig:hologic_results_vis}, we can notice consistent results with less noise, making this method the best choice for integration in the clinical practice.  

\begin{figure*}[t]
    \centering
    \includegraphics[width=0.65\textwidth]{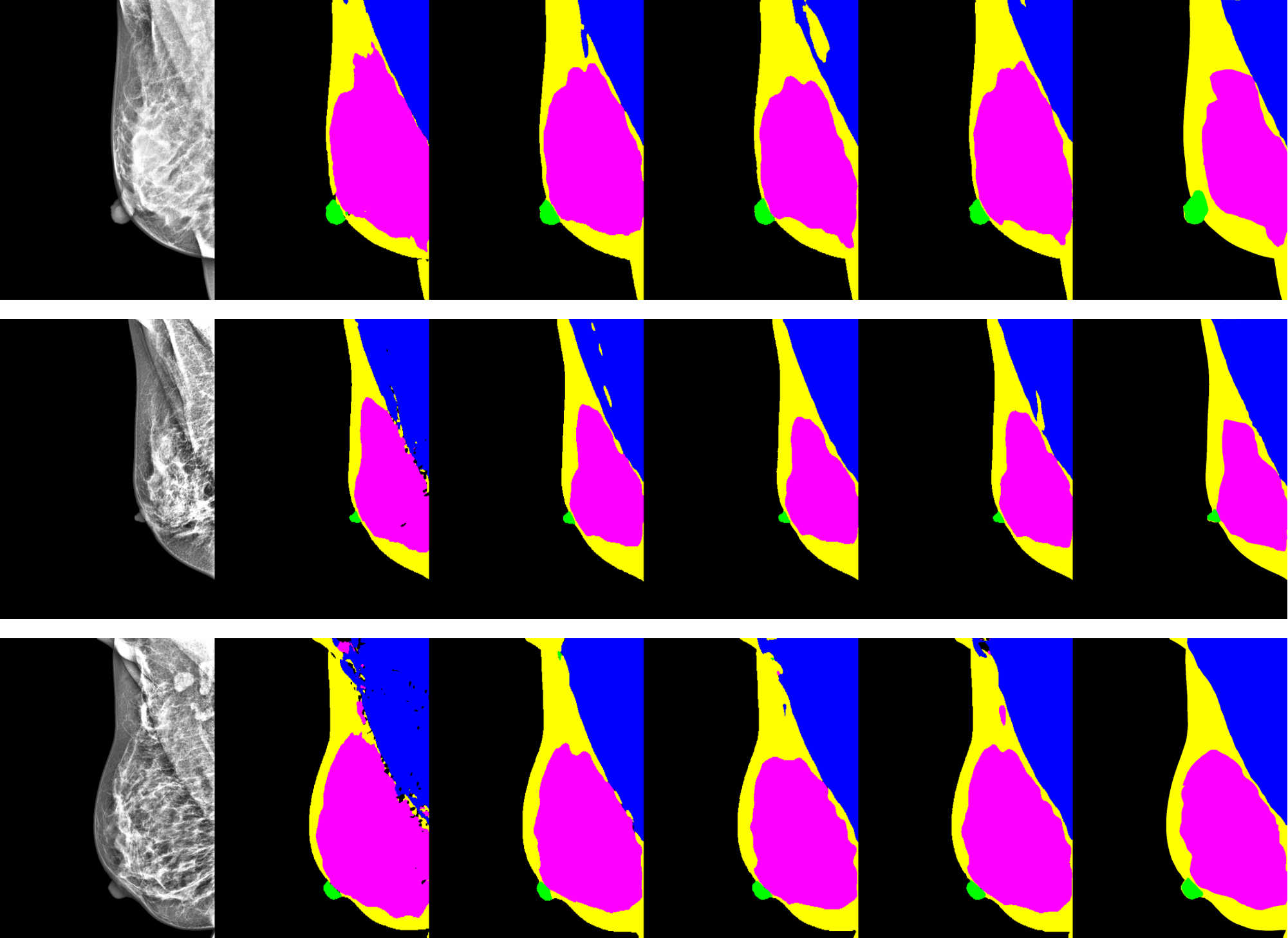}
    \caption{Visual results on the PLANMED dataset. Each row represents a different case. First column: input image. Second column: baseline result. Third column: image manipulation result. Fourth column: style transfer result. Fifth column: image manipulation and style transfer combination result. Sixth column: ground-truth annotation.}
    \label{fig:planmed_results_vis}
\end{figure*}

\begin{table*}[t]    \caption{Numerical results on the HOLOGIC dataset}
    \tiny
    \begin{center}
    \begin{tabular}{ J{2cm} J{3cm} J{1.2cm} J{1.2cm} J{1.2cm} J{1.2cm} J{1.2cm}} 
    \hline
    \textbf{Metric} & \textbf{Method} & \textbf{Nipple} & \textbf{Pectoral} & \textbf{Fib. Tissue} & \textbf{Fat. Tissue} & \textbf{Mean}\\
    \hline
    \multirow{4}{*}{Precision} & Baseline & 0.4805 & \textbf{0.9963} & 0.8664 & 0.7193 & 0.7656 \\
    & Image manipulation & 0.6903 & 0.9786 & 0.8986 & 0.8425 & 0.8525 \\
    & Style transfer & 0.7126 & 0.9816 & \textbf{0.9230} & 0.8079 & 0.8563 \\
    & Combination & \textbf{0.7138} & 0.9846 & 0.8984 & \textbf{0.8488} & \textbf{0.8614} \\
    \hline
    \multirow{4}{*}{Recall} & Baseline & 0.1545 & 0.7705 & 0.7212 & 0.5058 & 0.5380 \\
    & Image manipulation & 0.8535 & \textbf{0.9584} & 0.9272 & 0.8306 & 0.8924 \\
    & Style transfer & 0.7060 & 0.9477 & 0.9062 & \textbf{0.8510} & 0.8527 \\
    & Combination & \textbf{0.8620} & 0.9545 & \textbf{0.9308} & 0.8258 & \textbf{0.8933} \\
    \hline
    \multirow{4}{*}{Accuracy} & Baseline & 0.9981 & 0.9785 & 0.9169 & 0.8994 & 0.9482 \\
    & Image manipulation & 0.9990 & 0.9946 & 0.9628 & 0.9529 & 0.9773 \\
    & Style transfer & 0.9988 & 0.9933 & \textbf{0.9633} & 0.9487 & 0.9760 \\
    & Combination & \textbf{0.9991} & \textbf{0.9947} & 0.9631 & \textbf{0.9532} & \textbf{0.9775} \\
    \hline
    \multirow{4}{*}{Dice} & Baseline & 0.2245 & 0.8467 & 0.7826 & 0.5850 & 0.6097 \\
    & Image manipulation & 0.7295 & 0.9673 & 0.9085 & 0.8283 & 0.8584 \\
    & Style transfer & 0.6647 & 0.9621 & \textbf{0.9107} & 0.8212 & 0.8397 \\
    & Combination & \textbf{0.7501} & \textbf{0.9685} & 0.9103 & \textbf{0.8287} & \textbf{0.8644} \\
    \hline
    \multirow{4}{*}{IoU} & Baseline & 0.1463 & 0.7677 & 0.6487 & 0.4192 & 0.4955 \\
    & Image manipulation & 0.5901 & 0.9377 & 0.8349 & 0.7105 & 0.7683 \\
    & Style transfer & 0.5246 & 0.9302 & \textbf{0.8376} & 0.7007 & 0.7483 \\
    & Combination & \textbf{0.6174} & \textbf{0.9398} & 0.8375 & \textbf{0.7111} & \textbf{0.7764} \\
    \hline
    \multirow{4}{*}{Hausdorff} & Baseline & 0.0454 & 0.0298 & 0.0326 & 0.0314 & 0.0347 \\
    & Image manipulation & \textbf{0.0049} & 0.0168 & 0.0180 & \textbf{0.0211} & 0.0152 \\
    & Style transfer & 0.0398 & 0.0181 & \textbf{0.0174} & 0.0348 & 0.0275 \\
    & Combination & 0.0050 & \textbf{0.0163} & 0.0175 & 0.0213 & \textbf{0.0150} \\
    \hline
    \end{tabular}
    \end{center}
    \label{tab:numerical_results_hologic}
\end{table*}

\begin{table*}[t]    \caption{Statistical significance of the difference in results on the HOLOGIC dataset}
    \tiny
    \begin{center}
    \begin{tabular}{ J{1cm} J{1cm} J{13cm}} 
    \hline
    \textbf{Metric} & \textbf{Structure} & \textbf{Pairs} \\
    \hline
    Precision & Pectoral & (Baseline, Image manipulation), (Baseline, Style transfer), (Baseline, Combination) \\ \hline
    Precision & Fat. Tissue & (Baseline, Image manipulation), (Baseline, Style transfer), (Baseline, Combination) \\ \hline
    Precision & Mean & (Baseline, Image manipulation), (Baseline, Style transfer), (Baseline, Combination) \\ \hline
    Recall & Nipple & (Baseline, Image manipulation), (Baseline, Style transfer), (Baseline, Combination), (Image manipulation, Style transfer), (Style transfer, Combination) \\ \hline
    Recall & Pectoral & (Baseline, Image manipulation), (Baseline, Style transfer), (Baseline, Combination) \\ \hline
    Recall & Fib. Tissue & (Baseline, Image manipulation), (Baseline, Style transfer), (Baseline, Combination) \\ \hline
    Recall & Fat. Tissue & (Baseline, Image manipulation), (Baseline, Style transfer), (Baseline, Combination) \\ \hline
    Recall & Mean & (Baseline, Image manipulation), (Baseline, Style transfer), (Baseline, Combination) \\ \hline
    Accuracy & Nipple & (Baseline, Image manipulation), (Baseline, Style transfer), (Baseline, Combination) \\ \hline
    Accuracy & Pectoral & (Baseline, Image manipulation), (Baseline, Style transfer), (Baseline, Combination) \\ \hline
    Accuracy & Fib. Tissue & (Baseline, Image manipulation), (Baseline, Style transfer), (Baseline, Combination) \\ \hline
    Accuracy & Fat. Tissue & (Baseline, Image manipulation), (Baseline, Style transfer), (Baseline, Combination) \\ \hline
    Accuracy & Mean & (Baseline, Image manipulation), (Baseline, Style transfer), (Baseline, Combination) \\ \hline
    Dice & Nipple & (Baseline, Image manipulation), (Baseline, Style transfer), (Baseline, Combination) \\ \hline
    Dice & Pectoral & (Baseline, Image manipulation), (Baseline, Style transfer), (Baseline, Combination) \\ \hline
    Dice & Fib. Tissue & (Baseline, Image manipulation), (Baseline, Style transfer), (Baseline, Combination) \\ \hline
    Dice & Fat. Tissue & (Baseline, Image manipulation), (Baseline, Style transfer), (Baseline, Combination) \\ \hline
    Dice & Mean & (Baseline, Image manipulation), (Baseline, Style transfer), (Baseline, Combination) \\ \hline
    IoU & Nipple & (Baseline, Image manipulation), (Baseline, Style transfer), (Baseline, Combination) \\ \hline
    IoU & Pectoral & (Baseline, Image manipulation), (Baseline, Style transfer), (Baseline, Combination) \\ \hline
    IoU & Fib. Tissue & (Baseline, Image manipulation), (Baseline, Style transfer), (Baseline, Combination) \\ \hline
    IoU & Fat. Tissue & (Baseline, Image manipulation), (Baseline, Style transfer), (Baseline, Combination) \\ \hline
    IoU & Mean & (Baseline, Image manipulation), (Baseline, Style transfer), (Baseline, Combination) \\ \hline
    Hausdorff & Nipple & (Baseline, Image manipulation), (Baseline, Combination), (Image manipulation, Style transfer), (Style transfer, Combination) \\ \hline
    Hausdorff & Pectoral & (Baseline, Image manipulation), (Baseline, Style transfer), (Baseline, Combination) \\ \hline
    Hausdorff & Fib. Tissue & (Baseline, Image manipulation), (Baseline, Style transfer), (Baseline, Combination) \\ \hline
    Hausdorff & Fat. Tissue & (Baseline, Image manipulation), (Baseline, Combination), (Image manipulation, Style transfer), (Style transfer, Combination) \\ \hline
    Hausdorff & Mean & (Baseline, Image manipulation), (Baseline, Combination), (Image manipulation, Style transfer), (Style transfer, Combination) \\ \hline
    \end{tabular}
    \end{center}
    \label{tab:numerical_results_hologic_ss}
\end{table*}

\begin{figure*}[t]
    \centering
    \includegraphics[width=0.65\textwidth]{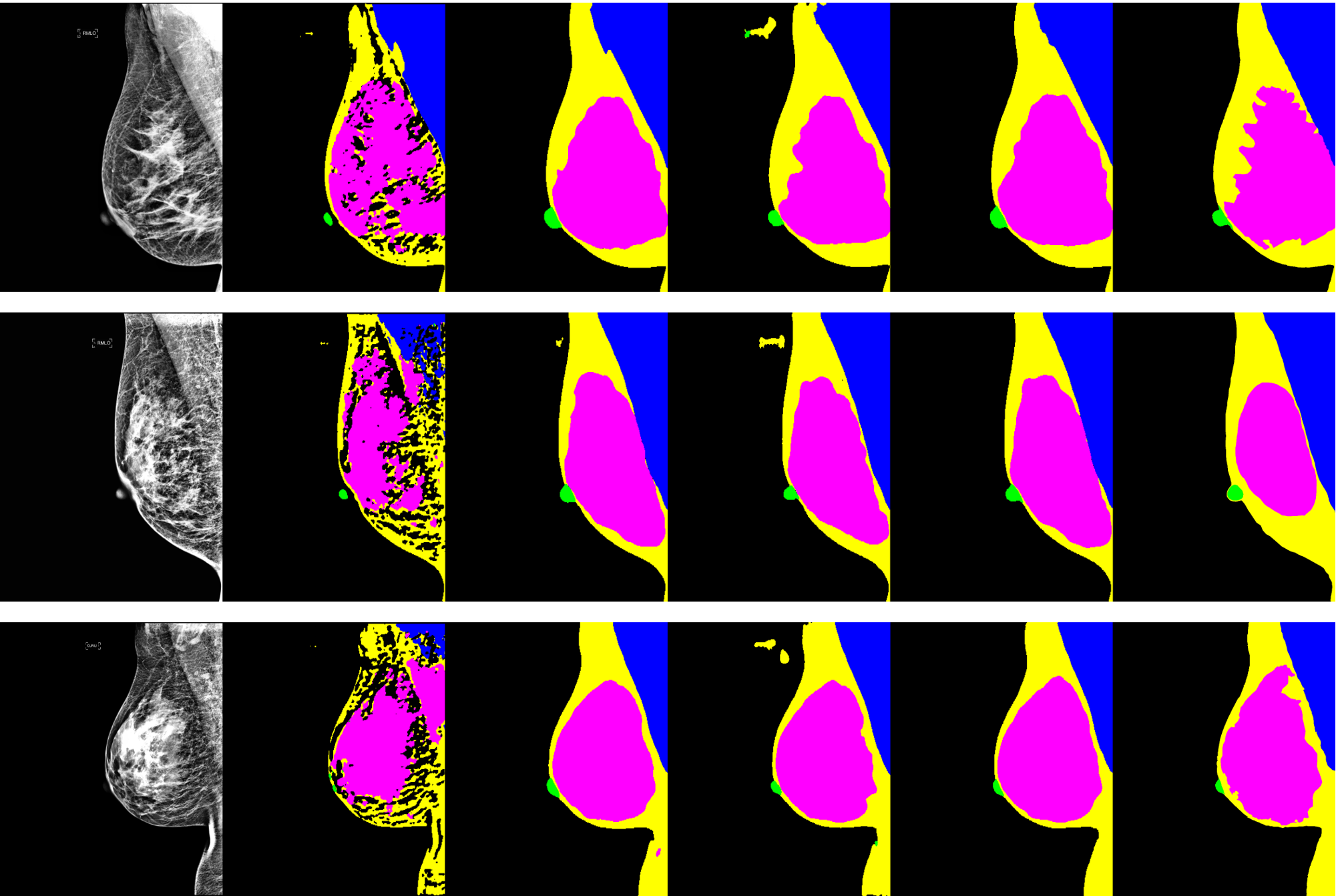}
    \caption{Visual results on the HOLOGIC dataset. Each row represents a different case. First column: input image. Second column: baseline result. Third column: image manipulation result. Fourth column: style transfer result. Fifth column: image manipulation and style transfer combination result. Sixth column: ground-truth annotation.}
    \label{fig:hologic_results_vis}
\end{figure*}

Figure~\ref{fig:results_uncertainty_vis} shows the uncertainty maps of the three proposed methods and the baseline method on four images from the HOLOGIC dataset. Notice how the proposed strategies minimize the high uncertainty regions compared to the baseline, concentrating them at the prediction boundaries. This suggests that the proposed models exhibit greater confidence and reliability when analyzing this type of image. Similar outcomes are observed when processing IMS and PLANMED images.

\begin{figure}[t]
    \centering
    \includegraphics[width=0.8\columnwidth]{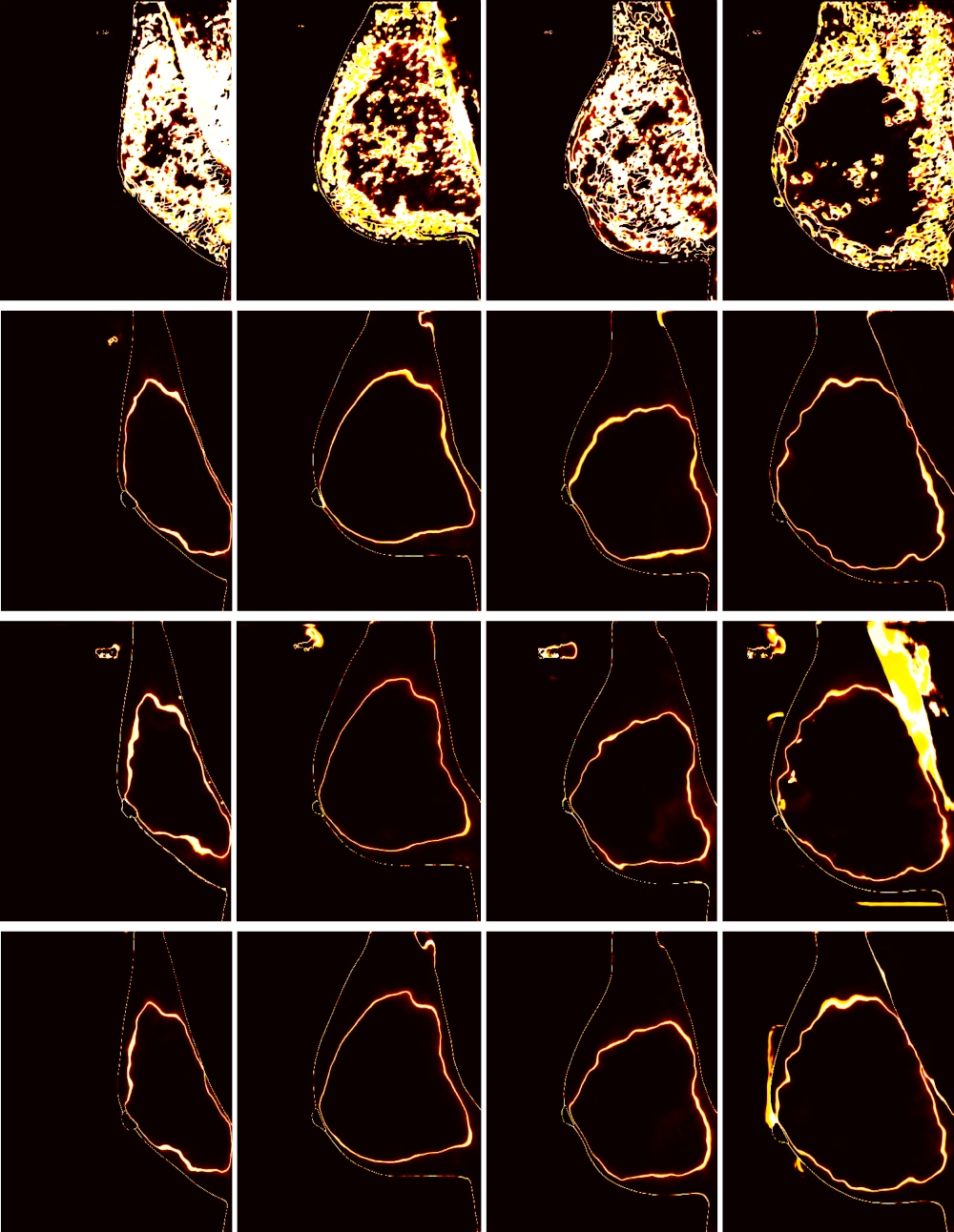}
    \caption{Uncertainty maps. Each column represents a different case. First row: baseline model. Second row: image manipulation. Third row: style transfer. Fourth row: image manipulation and style transfer combination.}
    \label{fig:results_uncertainty_vis}
\end{figure}

The presented approaches focus on digital mammography images, which represent the prevailing technology in contemporary practice. However, screen-film mammography remains in use across numerous medical centers, with extensive repositories established around this technology. The screen-film mammography images exhibit significant differences compared to digital mammography images, presenting challenges in their processing with the proposed models. Figure~\ref{fig:ddsm_results_vis} presents the predictions of the baseline and the combination methods on screen-film mammography images from the DDSM dataset \cite{heath1998current}. We can see how both methods present limitations when processing these images. However, the combination method seems to be more stable and robust.

\begin{figure}[t]
    \centering
    \includegraphics[width=0.8\columnwidth]{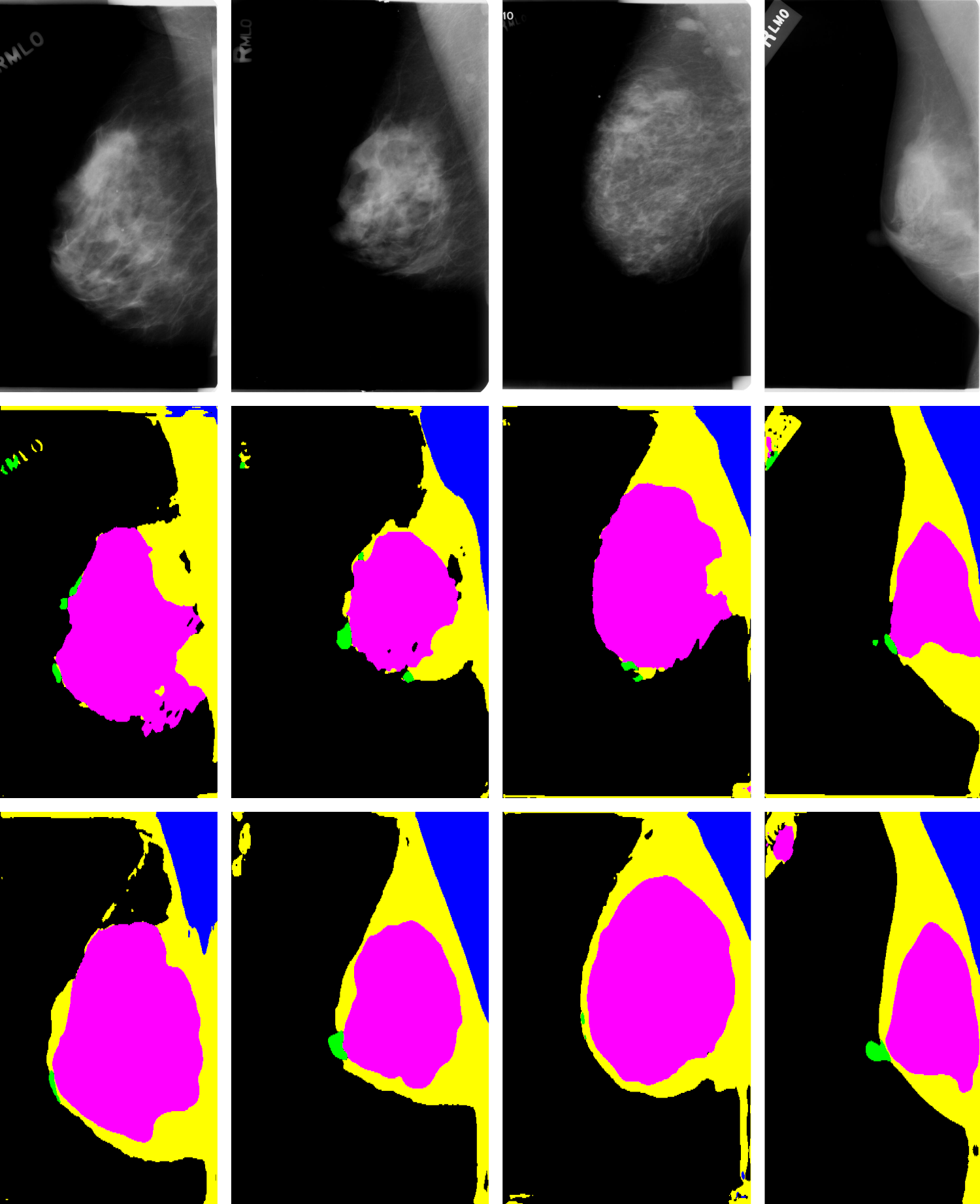}
    \caption{Results on screen-film mammography images from the DDSM dataset. Each column represents a different case. First row: input image. Second row: baseline. Third row: image manipulation and style transfer combination.}
    \label{fig:ddsm_results_vis}
\end{figure}

In \cite{verboom2024deep}, a deep learning-based approach is introduced for segmenting the pectoral muscle and breast in mammography images. The authors utilize a diverse dataset comprising mammography images from various vendor equipment, coupled with an aggressive augmentation procedure, to enhance generalization performance. In contrast to this approach, we focus on a dataset exclusively comprising GE images and extend the segmentation task to include additional structures of interest. Nonetheless, a comparative analysis can be conducted for the segmentation of the pectoral muscle on an unseen dataset, such as the HOLOGIC dataset employed in our experiments. Table~\ref{tab:pectoral_segmentation_comparison} presents the numerical results on the pectoral muscle segmentation task, comparing the performance of both \cite{verboom2024deep} and our combination method. Notice how our method achieves significantly superior metric values, demonstrating that it is more robust and confident for this task. Further, Figure~\ref{fig:pectoral_segmentation_comparison} shows some visual results, where we can see that our method is more consistent in predicting a single compact shape for the pectoral muscle, while the other presents noisy and incomplete predictions.

\begin{table}[t]    \caption{Numerical results for pectoral muscle segmentation on the HOLOGIC dataset}
    \tiny
    \begin{tabular}{ J{1.7cm} J{1.3cm} J{1.3cm} J{1.3cm} } 
    \hline
    \textbf{Method} & \textbf{Dice} & \textbf{IoU} & \textbf{Hausdorff} \\
    \hline
    \cite{verboom2024deep} & 0.8822 & 0.8145 & 0.0301 \\
    Ours & \textbf{0.9685} & \textbf{0.9398} & \textbf{0.0163} \\
    \hline
    \end{tabular}
    \label{tab:pectoral_segmentation_comparison}
\end{table}

\begin{figure}[t]
    \centering
    \includegraphics[width=0.8\columnwidth]{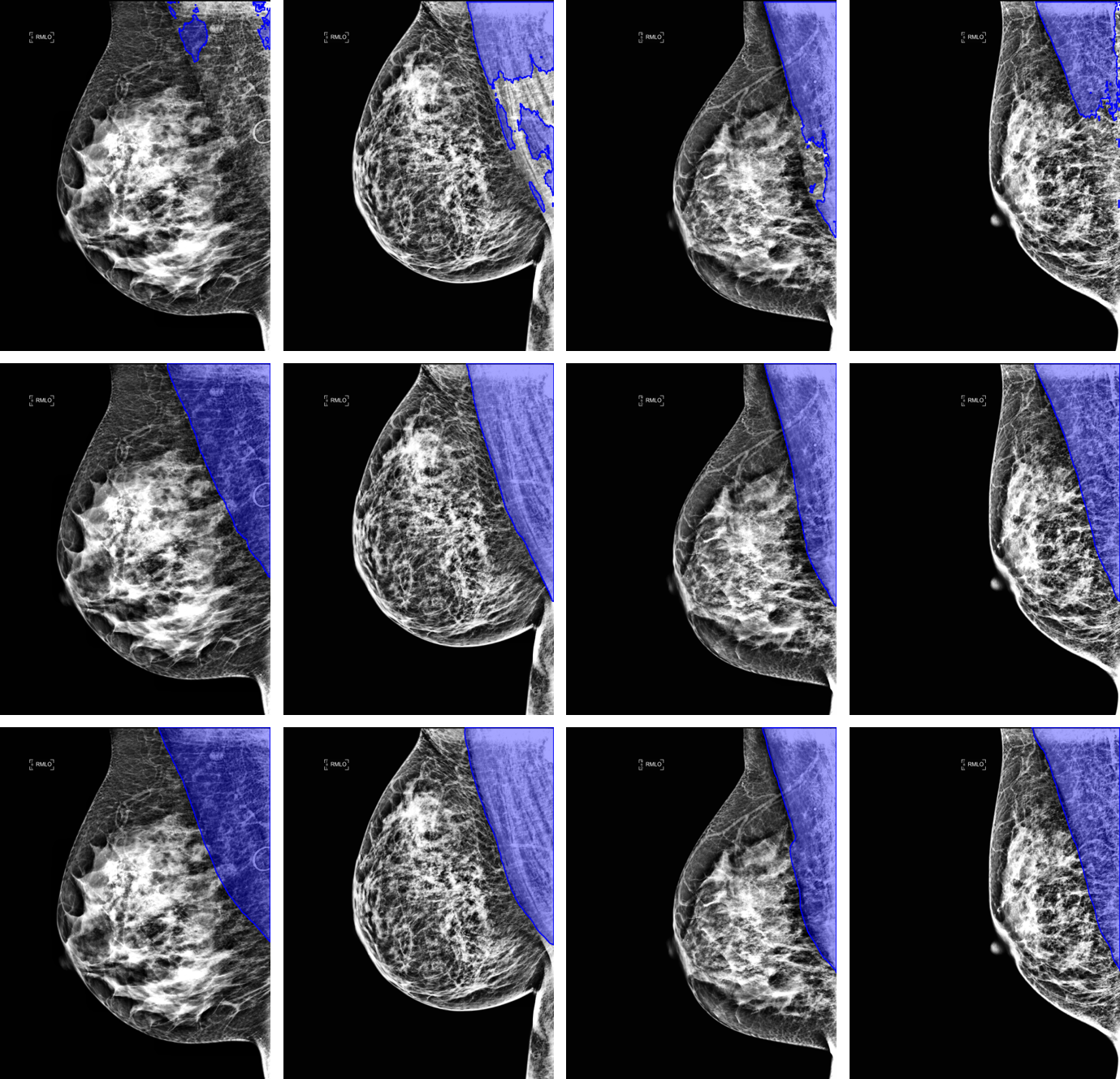}
    \caption{Pectoral segmentation comparison on the HOLOGIC dataset. Each column represents a different case. First row: results of the trained model introduced in \cite{verboom2024deep}. Second row: results of the proposed combination method. Third row: ground-truth annotation.}
    \label{fig:pectoral_segmentation_comparison}
\end{figure}

Although the presented experiments focus on the MLO view, our methods can be easily extended for the CC view. To show this adaptability, we selected the CC view mammography segmentation dataset introduced in \cite{sierra2024towards} for training and evaluation. This dataset consists of 5137 fully annotated GE images, where 3737, 943, and 457 images are considered for the training, validation, and test sets, respectively. Additionally, for the generalization evaluation, we consider a test set that consists a set of 34 fully-annotated HOLOGIC images. In both datasets, the same structures of interest presented in the previous sections are considered. 

We train a baseline model on the GE images using the same settings considered for the MLO view segmentation training. Then, leveraging our image manipulation method, we train another model using the same training settings as the CC view baseline model. Tables~\ref{tab:numerical_results_ge_cc} and ~\ref{tab:numerical_results_hologic_cc} present the numerical results on the GE and HOLOGIC test sets, respectively. Similarly to the behavior noticed for the MLO view, our method presents better results on the CC view HOLOGIC images while preserving the performance on the CC view GE images. Figure~\ref{fig:hologic_results_vis_cc} shows some visual results on HOLOGIC images, confirming the superior performance of our method on the processing of CC view images, even in the challenging segmentation of the pectoral muscle \cite{silva2023data}. 

\begin{table*}[t]    \caption{Numerical results on CC view GE images}
    \tiny
    \begin{center}
    \begin{tabular}{ J{2cm} J{3cm} J{1.2cm} J{1.2cm} J{1.2cm} J{1.2cm} J{1.2cm}} 
    \hline
    \textbf{Metric} & \textbf{Method} & \textbf{Nipple} & \textbf{Pectoral} & \textbf{Fib. Tissue} & \textbf{Fat. Tissue} & \textbf{Mean}\\
    \hline
    \multirow{2}{*}{Dice} & Baseline & \textbf{0.8610} & \textbf{0.3998} & \textbf{0.9569} & \textbf{0.9103} & \textbf{0.7820} \\
    & Image manipulation & 0.8473 & 0.3870 & 0.9543 & 0.9083 & 0.7742 \\
    \hline
    \multirow{2}{*}{IoU} & Baseline & \textbf{0.7731} & 0.8289 & \textbf{0.9194} & \textbf{0.8417} & \textbf{0.8408} \\
    & Image manipulation & 0.7567 & \textbf{0.8456} & 0.9151 & 0.8380 & 0.8389 \\
    \hline
    \multirow{2}{*}{Hausdorff} & Baseline & \textbf{0.0003} & \textbf{0.0008} & \textbf{0.0022} & \textbf{0.0023} & \textbf{0.0015} \\
    & Image manipulation & 0.0003 & 0.0009 & 0.0022 & 0.0024 & 0.0016 \\
    \hline
    \end{tabular}
    \end{center}
    \label{tab:numerical_results_ge_cc}
\end{table*}

\begin{table*}[t]    \caption{Numerical results on CC view HOLOGIC images}
    \tiny
    \begin{center}
    \begin{tabular}{ J{2cm} J{3cm} J{1.2cm} J{1.2cm} J{1.2cm} J{1.2cm} J{1.2cm}} 
    \hline
    \textbf{Metric} & \textbf{Method} & \textbf{Nipple} & \textbf{Pectoral} & \textbf{Fib. Tissue} & \textbf{Fat. Tissue} & \textbf{Mean}\\
    \hline
    \multirow{2}{*}{Dice} & Baseline & 0.1632 & 0.3059 & 0.8259 & 0.6608 & 0.4889 \\
    & Image manipulation & \textbf{0.6990} & \textbf{0.4136} & \textbf{0.8767} & \textbf{0.8108} & \textbf{0.7000} \\
    \hline
    \multirow{2}{*}{IoU} & Baseline & 0.1011 & 0.5331 & 0.7070 & 0.4949 & 0.4590 \\
    & Image manipulation & \textbf{0.5613} & \textbf{0.7126} & \textbf{0.7865} & \textbf{0.6850} & \textbf{0.6863} \\
    \hline
    \multirow{2}{*}{Hausdorff} & Baseline & 0.0500 & 0.0117 & 0.0262 & 0.0356 & 0.0341 \\
    & Image manipulation & \textbf{0.0032} & \textbf{0.0053} & \textbf{0.0250} & \textbf{0.0247} & \textbf{0.0167} \\
    \hline
    \end{tabular}
    \end{center}
    \label{tab:numerical_results_hologic_cc}
\end{table*}

\begin{figure*}[t]
    \centering
    \includegraphics[width=0.65\textwidth]{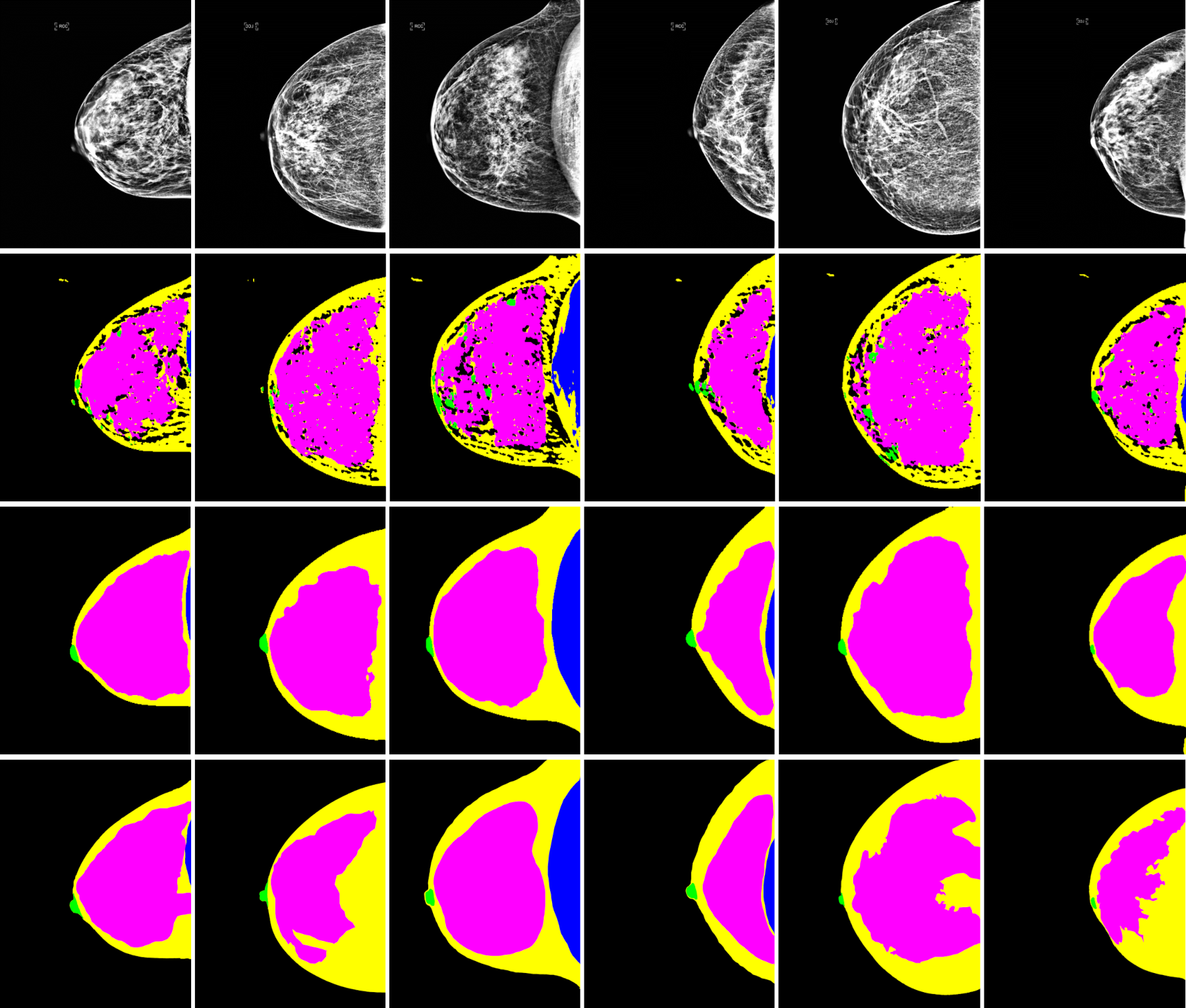}
    \caption{Visual results on CC view HOLOGIC images. Each column represents a different case. First row: input image. Second row: baseline result. Third row: image manipulation result. Fourth row: ground-truth annotation.}
    \label{fig:hologic_results_vis_cc}
\end{figure*}
\section{Discussion}
\label{sec:discussion}

The numerical results demonstrate the effectiveness of our proposed data-centric augmentation strategies in improving the generalization of deep learning models for mammography image segmentation. Compared to the baseline method, our approach significantly enhances segmentation performance across images from different vendor equipment. These findings are further validated through the visualization of predictions and corresponding uncertainty maps.

Although our evaluations are based on a limited number of different vendor equipment, the corresponding images represent the most diverse samples compared to those used for training, i.e. GE images. We expect the trained models to perform even better on images that closely resemble the GE images, such as those generated by Siemens or Fujifilm equipment.

We present visual results on screen-film mammography images, demonstrating a subtle enhancement achieved by the proposed method compared to the baseline. However, the predictions include noisy structures, requiring further post-processing operations to achieve a reliable segmentation. Further exploration of generalization across this domain and other image settings remains an open problem that we plan to address in future work.

Our assessment of CC view images demonstrates the applicability of the image intensity manipulation method to this domain. We expect that the style transfer method and the combination of both will exhibit similar efficacy. However, further investigation is needed, including an evaluation of pectoral muscle detection, which is challenging in the CC view.

\section{Conclusion}
\label{sec:conclusion}

We address the challenge of segmenting landmark structures in mammography images, which is crucial for breast cancer assessment. Our approach considers data-centric strategies to enrich training data for deep learning-based segmentation. This involves augmenting training samples through annotation-guided image intensity manipulation and style transfer to improve generalization beyond conventional training methods.

Our findings demonstrate the effectiveness of the proposed methods in achieving improved generalization across various vendor equipment, even when considering training data from a single vendor. This approach avoids the need to generate new training images and manual annotations, thus reducing labor costs and saving time in clinical settings.

While we highlighted the importance of segmenting landmark structures for assessing cancer risk and image acquisition adequacy, our experiments do not directly evaluate the efficacy of the proposed methods for these tasks. In future work, we aim to explore these applications and others using them.
\fi

\if\isanonymized1
\else

\section*{Statements and declarations}

\subsection*{Funding} 
The authors declare that no funds, grants, or other support were received during the preparation of this manuscript.

\subsection*{Competing Interests} 
The authors declare that they have no conflict of interest.

\subsection*{Author Contributions}
\textbf{Jan Hurtado and João P. Maia:} study conception and design, development and evaluation of the proposed methods, and manuscript preparation and analysis. \textbf{Cesar A. Sierra Franco and Alberto Raposo:} study conception and design, and manuscript preparation and analysis.

\subsection*{Ethics approval} 
Not applicable. This study uses publicly available and private data from previously published research, which had obtained the necessary ethical approvals.

\subsection*{Consent to participate}
Not applicable. This study does not involve new data collection or human participants. The original study obtained consent as required.

\subsection*{Consent to publish}
Not applicable. This study uses data from previously published research, and no new individual data is disclosed.

\fi

\if\istitlepage1
\else
\bibliography{main}


\begin{thebibliography}{32}
\ifx \bisbn   \undefined \def \bisbn  #1{ISBN #1}\fi
\ifx \binits  \undefined \def \binits#1{#1}\fi
\ifx \bauthor  \undefined \def \bauthor#1{#1}\fi
\ifx \batitle  \undefined \def \batitle#1{#1}\fi
\ifx \bjtitle  \undefined \def \bjtitle#1{#1}\fi
\ifx \bvolume  \undefined \def \bvolume#1{\textbf{#1}}\fi
\ifx \byear  \undefined \def \byear#1{#1}\fi
\ifx \bissue  \undefined \def \bissue#1{#1}\fi
\ifx \bfpage  \undefined \def \bfpage#1{#1}\fi
\ifx \blpage  \undefined \def \blpage #1{#1}\fi
\ifx \burl  \undefined \def \burl#1{\textsf{#1}}\fi
\ifx \doiurl  \undefined \def \doiurl#1{\url{https://doi.org/#1}}\fi
\ifx \betal  \undefined \def \betal{\textit{et al.}}\fi
\ifx \binstitute  \undefined \def \binstitute#1{#1}\fi
\ifx \binstitutionaled  \undefined \def \binstitutionaled#1{#1}\fi
\ifx \bctitle  \undefined \def \bctitle#1{#1}\fi
\ifx \beditor  \undefined \def \beditor#1{#1}\fi
\ifx \bpublisher  \undefined \def \bpublisher#1{#1}\fi
\ifx \bbtitle  \undefined \def \bbtitle#1{#1}\fi
\ifx \bedition  \undefined \def \bedition#1{#1}\fi
\ifx \bseriesno  \undefined \def \bseriesno#1{#1}\fi
\ifx \blocation  \undefined \def \blocation#1{#1}\fi
\ifx \bsertitle  \undefined \def \bsertitle#1{#1}\fi
\ifx \bsnm \undefined \def \bsnm#1{#1}\fi
\ifx \bsuffix \undefined \def \bsuffix#1{#1}\fi
\ifx \bparticle \undefined \def \bparticle#1{#1}\fi
\ifx \barticle \undefined \def \barticle#1{#1}\fi
\bibcommenthead
\ifx \bconfdate \undefined \def \bconfdate #1{#1}\fi
\ifx \botherref \undefined \def \botherref #1{#1}\fi
\ifx \url \undefined \def \url#1{\textsf{#1}}\fi
\ifx \bchapter \undefined \def \bchapter#1{#1}\fi
\ifx \bbook \undefined \def \bbook#1{#1}\fi
\ifx \bcomment \undefined \def \bcomment#1{#1}\fi
\ifx \oauthor \undefined \def \oauthor#1{#1}\fi
\ifx \citeauthoryear \undefined \def \citeauthoryear#1{#1}\fi
\ifx \endbibitem  \undefined \def \endbibitem {}\fi
\ifx \bconflocation  \undefined \def \bconflocation#1{#1}\fi
\ifx \arxivurl  \undefined \def \arxivurl#1{\textsf{#1}}\fi
\csname PreBibitemsHook\endcsname

\bibitem[\protect\citeauthoryear{Sierra-Franco et~al.}{2024}]{sierra2024towards}
\begin{botherref}
\oauthor{\bsnm{Sierra-Franco}, \binits{C.A.}},
\oauthor{\bsnm{Hurtado}, \binits{J.}},
\oauthor{\bsnm{A~Thomaz}, \binits{V.}},
\oauthor{\bsnm{Cruz}, \binits{L.C.}},
\oauthor{\bsnm{Silva}, \binits{S.V.}},
\oauthor{\bsnm{Silva-Calpa}, \binits{G.F.M.}},
\oauthor{\bsnm{Raposo}, \binits{A.}}:
Towards automated semantic segmentation in mammography images for enhanced clinical applications.
Journal of Imaging Informatics in Medicine,
1--21
(2024)
\end{botherref}
\endbibitem

\bibitem[\protect\citeauthoryear{Rampun et~al.}{2019}]{rampun2019breast}
\begin{barticle}
\bauthor{\bsnm{Rampun}, \binits{A.}},
\bauthor{\bsnm{L{\'o}pez-Linares}, \binits{K.}},
\bauthor{\bsnm{Morrow}, \binits{P.J.}},
\bauthor{\bsnm{Scotney}, \binits{B.W.}},
\bauthor{\bsnm{Wang}, \binits{H.}},
\bauthor{\bsnm{Oca{\~n}a}, \binits{I.G.}},
\bauthor{\bsnm{Maclair}, \binits{G.}},
\bauthor{\bsnm{Zwiggelaar}, \binits{R.}},
\bauthor{\bsnm{Ballester}, \binits{M.A.G.}},
\bauthor{\bsnm{Mac{\'\i}a}, \binits{I.}}:
\batitle{Breast pectoral muscle segmentation in mammograms using a modified holistically-nested edge detection network}.
\bjtitle{Medical image analysis}
\bvolume{57},
\bfpage{1}--\blpage{17}
(\byear{2019})
\end{barticle}
\endbibitem

\bibitem[\protect\citeauthoryear{Soleimani and Michailovich}{2020}]{soleimani2020segmentation}
\begin{barticle}
\bauthor{\bsnm{Soleimani}, \binits{H.}},
\bauthor{\bsnm{Michailovich}, \binits{O.V.}}:
\batitle{On segmentation of pectoral muscle in digital mammograms by means of deep learning}.
\bjtitle{IEEE Access}
\bvolume{8},
\bfpage{204173}--\blpage{204182}
(\byear{2020})
\end{barticle}
\endbibitem

\bibitem[\protect\citeauthoryear{Ali et~al.}{2020}]{ali2020enhancing}
\begin{barticle}
\bauthor{\bsnm{Ali}, \binits{M.J.}},
\bauthor{\bsnm{Raza}, \binits{B.}},
\bauthor{\bsnm{Shahid}, \binits{A.R.}},
\bauthor{\bsnm{Mahmood}, \binits{F.}},
\bauthor{\bsnm{Yousuf}, \binits{M.A.}},
\bauthor{\bsnm{Dar}, \binits{A.H.}},
\bauthor{\bsnm{Iqbal}, \binits{U.}}:
\batitle{Enhancing breast pectoral muscle segmentation performance by using skip connections in fully convolutional network}.
\bjtitle{International Journal of Imaging Systems and Technology}
\bvolume{30}(\bissue{4}),
\bfpage{1108}--\blpage{1118}
(\byear{2020})
\end{barticle}
\endbibitem

\bibitem[\protect\citeauthoryear{Rubio and Montiel}{2021}]{rubio2021multicriteria}
\begin{barticle}
\bauthor{\bsnm{Rubio}, \binits{Y.}},
\bauthor{\bsnm{Montiel}, \binits{O.}}:
\batitle{Multicriteria evaluation of deep neural networks for semantic segmentation of mammographies}.
\bjtitle{Axioms}
\bvolume{10}(\bissue{3}),
\bfpage{180}
(\byear{2021})
\end{barticle}
\endbibitem

\bibitem[\protect\citeauthoryear{Guo et~al.}{2020}]{guo2020automatic}
\begin{barticle}
\bauthor{\bsnm{Guo}, \binits{Y.}},
\bauthor{\bsnm{Zhao}, \binits{W.}},
\bauthor{\bsnm{Li}, \binits{S.}},
\bauthor{\bsnm{Zhang}, \binits{Y.}},
\bauthor{\bsnm{Lu}, \binits{Y.}}:
\batitle{Automatic segmentation of the pectoral muscle based on boundary identification and shape prediction}.
\bjtitle{Physics in Medicine \& Biology}
\bvolume{65}(\bissue{4}),
\bfpage{045016}
(\byear{2020})
\end{barticle}
\endbibitem

\bibitem[\protect\citeauthoryear{Yu et~al.}{2022}]{yu2022pemnet}
\begin{barticle}
\bauthor{\bsnm{Yu}, \binits{X.}},
\bauthor{\bsnm{Wang}, \binits{S.-H.}},
\bauthor{\bsnm{G{\'o}rriz}, \binits{J.M.}},
\bauthor{\bsnm{Jiang}, \binits{X.-W.}},
\bauthor{\bsnm{Guttery}, \binits{D.S.}},
\bauthor{\bsnm{Zhang}, \binits{Y.-D.}}:
\batitle{Pemnet for pectoral muscle segmentation}.
\bjtitle{Biology}
\bvolume{11}(\bissue{1}),
\bfpage{134}
(\byear{2022})
\end{barticle}
\endbibitem

\bibitem[\protect\citeauthoryear{Verboom et~al.}{2024}]{verboom2024deep}
\begin{barticle}
\bauthor{\bsnm{Verboom}, \binits{S.D.}},
\bauthor{\bsnm{Caballo}, \binits{M.}},
\bauthor{\bsnm{Peters}, \binits{J.}},
\bauthor{\bsnm{Gommers}, \binits{J.}},
\bauthor{\bsnm{Oever}, \binits{D.}},
\bauthor{\bsnm{Broeders}, \binits{M.J.}},
\bauthor{\bsnm{Teuwen}, \binits{J.}},
\bauthor{\bsnm{Sechopoulos}, \binits{I.}}:
\batitle{Deep learning-based breast region segmentation in raw and processed digital mammograms: generalization across views and vendors}.
\bjtitle{Journal of Medical Imaging}
\bvolume{11}(\bissue{1}),
\bfpage{014001}--\blpage{014001}
(\byear{2024})
\end{barticle}
\endbibitem

\bibitem[\protect\citeauthoryear{Yin et~al.}{1994}]{yin1994computerized}
\begin{barticle}
\bauthor{\bsnm{Yin}, \binits{F.-F.}},
\bauthor{\bsnm{Giger}, \binits{M.L.}},
\bauthor{\bsnm{Doi}, \binits{K.}},
\bauthor{\bsnm{Vyborny}, \binits{C.J.}},
\bauthor{\bsnm{Schmidt}, \binits{R.A.}}:
\batitle{Computerized detection of masses in digital mammograms: Automated alignment of breast images and its effect on bilateral-subtraction technique}.
\bjtitle{Medical Physics}
\bvolume{21}(\bissue{3}),
\bfpage{445}--\blpage{452}
(\byear{1994})
\end{barticle}
\endbibitem

\bibitem[\protect\citeauthoryear{M{\'e}ndez et~al.}{1996}]{mendez1996automatic}
\begin{barticle}
\bauthor{\bsnm{M{\'e}ndez}, \binits{A.J.}},
\bauthor{\bsnm{Tahoces}, \binits{P.G.}},
\bauthor{\bsnm{Lado}, \binits{M.J.}},
\bauthor{\bsnm{Souto}, \binits{M.}},
\bauthor{\bsnm{Correa}, \binits{J.}},
\bauthor{\bsnm{Vidal}, \binits{J.J.}}:
\batitle{Automatic detection of breast border and nipple in digital mammograms}.
\bjtitle{Computer methods and programs in biomedicine}
\bvolume{49}(\bissue{3}),
\bfpage{253}--\blpage{262}
(\byear{1996})
\end{barticle}
\endbibitem

\bibitem[\protect\citeauthoryear{Chandrasekhar and Attikiouzel}{1997}]{chandrasekhar1997simple}
\begin{barticle}
\bauthor{\bsnm{Chandrasekhar}, \binits{R.}},
\bauthor{\bsnm{Attikiouzel}, \binits{Y.}}:
\batitle{A simple method for automatically locating the nipple on mammograms}.
\bjtitle{IEEE transactions on medical imaging}
\bvolume{16}(\bissue{5}),
\bfpage{483}--\blpage{494}
(\byear{1997})
\end{barticle}
\endbibitem

\bibitem[\protect\citeauthoryear{Mustra et~al.}{2009}]{mustra2009nipple}
\begin{bchapter}
\bauthor{\bsnm{Mustra}, \binits{M.}},
\bauthor{\bsnm{Bozek}, \binits{J.}},
\bauthor{\bsnm{Grgic}, \binits{M.}}:
\bctitle{Nipple detection in craniocaudal digital mammograms}.
In: \bbtitle{2009 International Symposium ELMAR},
pp. \bfpage{15}--\blpage{18}
(\byear{2009}).
\bcomment{IEEE}
\end{bchapter}
\endbibitem

\bibitem[\protect\citeauthoryear{Casti et~al.}{2013}]{casti2013automatic}
\begin{barticle}
\bauthor{\bsnm{Casti}, \binits{P.}},
\bauthor{\bsnm{Mencattini}, \binits{A.}},
\bauthor{\bsnm{Salmeri}, \binits{M.}},
\bauthor{\bsnm{Ancona}, \binits{A.}},
\bauthor{\bsnm{Mangieri}, \binits{F.F.}},
\bauthor{\bsnm{Pepe}, \binits{M.L.}},
\bauthor{\bsnm{Rangayyan}, \binits{R.M.}}:
\batitle{Automatic detection of the nipple in screen-film and full-field digital mammograms using a novel hessian-based method}.
\bjtitle{Journal of digital imaging}
\bvolume{26},
\bfpage{948}--\blpage{957}
(\byear{2013})
\end{barticle}
\endbibitem

\bibitem[\protect\citeauthoryear{Zhou et~al.}{2004}]{zhou2004computerized}
\begin{barticle}
\bauthor{\bsnm{Zhou}, \binits{C.}},
\bauthor{\bsnm{Chan}, \binits{H.-P.}},
\bauthor{\bsnm{Paramagul}, \binits{C.}},
\bauthor{\bsnm{Roubidoux}, \binits{M.A.}},
\bauthor{\bsnm{Sahiner}, \binits{B.}},
\bauthor{\bsnm{Hadjiiski}, \binits{L.M.}},
\bauthor{\bsnm{Petrick}, \binits{N.}}:
\batitle{Computerized nipple identification for multiple image analysis in computer-aided diagnosis: Computerized nipple identification on mammograms}.
\bjtitle{Medical Physics}
\bvolume{31}(\bissue{10}),
\bfpage{2871}--\blpage{2882}
(\byear{2004})
\end{barticle}
\endbibitem

\bibitem[\protect\citeauthoryear{Kinoshita et~al.}{2008}]{kinoshita2008radon}
\begin{barticle}
\bauthor{\bsnm{Kinoshita}, \binits{S.K.}},
\bauthor{\bsnm{Azevedo-Marques}, \binits{P.M.}},
\bauthor{\bsnm{Pereira}, \binits{R.R.}},
\bauthor{\bsnm{Rodrigues}, \binits{J.A.H.}},
\bauthor{\bsnm{Rangayyan}, \binits{R.M.}}:
\batitle{Radon-domain detection of the nipple and the pectoral muscle in mammograms}.
\bjtitle{Journal of digital imaging}
\bvolume{21},
\bfpage{37}--\blpage{49}
(\byear{2008})
\end{barticle}
\endbibitem

\bibitem[\protect\citeauthoryear{Jiang et~al.}{2019}]{jiang2019radiomic}
\begin{barticle}
\bauthor{\bsnm{Jiang}, \binits{J.}},
\bauthor{\bsnm{Zhang}, \binits{Y.}},
\bauthor{\bsnm{Lu}, \binits{Y.}},
\bauthor{\bsnm{Guo}, \binits{Y.}},
\bauthor{\bsnm{Chen}, \binits{H.}}:
\batitle{A radiomic feature--based nipple detection algorithm on digital mammography}.
\bjtitle{Medical physics}
\bvolume{46}(\bissue{10}),
\bfpage{4381}--\blpage{4391}
(\byear{2019})
\end{barticle}
\endbibitem

\bibitem[\protect\citeauthoryear{Lin et~al.}{2019}]{lin2019nipple}
\begin{bchapter}
\bauthor{\bsnm{Lin}, \binits{Y.}},
\bauthor{\bsnm{Li}, \binits{M.}},
\bauthor{\bsnm{Chen}, \binits{S.}},
\bauthor{\bsnm{Yu}, \binits{L.}},
\bauthor{\bsnm{Ma}, \binits{F.}}:
\bctitle{Nipple detection in mammogram using a new convolutional neural network architecture}.
In: \bbtitle{2019 12th International Congress on Image and Signal Processing, BioMedical Engineering and Informatics (CISP-BMEI)},
pp. \bfpage{1}--\blpage{6}
(\byear{2019}).
\bcomment{IEEE}
\end{bchapter}
\endbibitem

\bibitem[\protect\citeauthoryear{He et~al.}{2015}]{he2015review}
\begin{botherref}
\oauthor{\bsnm{He}, \binits{W.}},
\oauthor{\bsnm{Juette}, \binits{A.}},
\oauthor{\bsnm{Denton}, \binits{E.R.}},
\oauthor{\bsnm{Oliver}, \binits{A.}},
\oauthor{\bsnm{Mart{\'\i}}, \binits{R.}},
\oauthor{\bsnm{Zwiggelaar}, \binits{R.}}, et al.:
A review on automatic mammographic density and parenchymal segmentation.
International journal of breast cancer
\textbf{2015}
(2015)
\end{botherref}
\endbibitem

\bibitem[\protect\citeauthoryear{Matsubara et~al.}{2001}]{matsubara2001automated}
\begin{bchapter}
\bauthor{\bsnm{Matsubara}, \binits{T.}},
\bauthor{\bsnm{Yamazaki}, \binits{D.}},
\bauthor{\bsnm{Kato}, \binits{M.}},
\bauthor{\bsnm{Hara}, \binits{T.}},
\bauthor{\bsnm{Fujita}, \binits{H.}},
\bauthor{\bsnm{Iwase}, \binits{T.}},
\bauthor{\bsnm{Endo}, \binits{T.}}:
\bctitle{An automated classification scheme for mammograms based on amount and distribution of fibroglandular breast tissue density}.
In: \bbtitle{International Congress Series},
vol. \bseriesno{1230},
pp. \bfpage{545}--\blpage{552}
(\byear{2001}).
\bcomment{Elsevier}
\end{bchapter}
\endbibitem

\bibitem[\protect\citeauthoryear{El-Zaart}{2010}]{el2010expectation}
\begin{barticle}
\bauthor{\bsnm{El-Zaart}, \binits{A.}}:
\batitle{Expectation--maximization technique for fibro-glandular discs detection in mammography images}.
\bjtitle{Computers in Biology and Medicine}
\bvolume{40}(\bissue{4}),
\bfpage{392}--\blpage{401}
(\byear{2010})
\end{barticle}
\endbibitem

\bibitem[\protect\citeauthoryear{Torres et~al.}{2019}]{torres2019morphological}
\begin{bchapter}
\bauthor{\bsnm{Torres}, \binits{G.F.}},
\bauthor{\bsnm{Sassi}, \binits{A.}},
\bauthor{\bsnm{Arponen}, \binits{O.}},
\bauthor{\bsnm{Holli-Helenius}, \binits{K.}},
\bauthor{\bsnm{L{\"a}{\"a}peri}, \binits{A.-L.}},
\bauthor{\bsnm{Rinta-Kiikka}, \binits{I.}},
\bauthor{\bsnm{K{\"a}m{\"a}r{\"a}inen}, \binits{J.}},
\bauthor{\bsnm{Pertuz}, \binits{S.}}:
\bctitle{Morphological area gradient: System-independent dense tissue segmentation in mammography images}.
In: \bbtitle{2019 41st Annual International Conference of the IEEE Engineering in Medicine and Biology Society (EMBC)},
pp. \bfpage{4855}--\blpage{4858}
(\byear{2019}).
\bcomment{IEEE}
\end{bchapter}
\endbibitem

\bibitem[\protect\citeauthoryear{Saffari et~al.}{2020}]{saffari2020fully}
\begin{barticle}
\bauthor{\bsnm{Saffari}, \binits{N.}},
\bauthor{\bsnm{Rashwan}, \binits{H.A.}},
\bauthor{\bsnm{Abdel-Nasser}, \binits{M.}},
\bauthor{\bsnm{Kumar~Singh}, \binits{V.}},
\bauthor{\bsnm{Arenas}, \binits{M.}},
\bauthor{\bsnm{Mangina}, \binits{E.}},
\bauthor{\bsnm{Herrera}, \binits{B.}},
\bauthor{\bsnm{Puig}, \binits{D.}}:
\batitle{Fully automated breast density segmentation and classification using deep learning}.
\bjtitle{Diagnostics}
\bvolume{10}(\bissue{11}),
\bfpage{988}
(\byear{2020})
\end{barticle}
\endbibitem

\bibitem[\protect\citeauthoryear{Larroza et~al.}{2022}]{larroza2022breast}
\begin{barticle}
\bauthor{\bsnm{Larroza}, \binits{A.}},
\bauthor{\bsnm{P{\'e}rez-Benito}, \binits{F.J.}},
\bauthor{\bsnm{Perez-Cortes}, \binits{J.-C.}},
\bauthor{\bsnm{Rom{\'a}n}, \binits{M.}},
\bauthor{\bsnm{Poll{\'a}n}, \binits{M.}},
\bauthor{\bsnm{P{\'e}rez-G{\'o}mez}, \binits{B.}},
\bauthor{\bsnm{Salas-Trejo}, \binits{D.}},
\bauthor{\bsnm{Casals}, \binits{M.}},
\bauthor{\bsnm{Llobet}, \binits{R.}}:
\batitle{Breast dense tissue segmentation with noisy labels: A hybrid threshold-based and mask-based approach}.
\bjtitle{Diagnostics}
\bvolume{12}(\bissue{8}),
\bfpage{1822}
(\byear{2022})
\end{barticle}
\endbibitem

\bibitem[\protect\citeauthoryear{Hu et~al.}{2022}]{hu2022breast}
\begin{bchapter}
\bauthor{\bsnm{Hu}, \binits{J.}},
\bauthor{\bsnm{Liu}, \binits{Z.}},
\bauthor{\bsnm{Wang}, \binits{Q.}}:
\bctitle{Breast density segmentation in mammograms based on dual attention mechanism}.
In: \bbtitle{Proceedings of the 3rd International Symposium on Artificial Intelligence for Medicine Sciences},
pp. \bfpage{430}--\blpage{435}
(\byear{2022})
\end{bchapter}
\endbibitem

\bibitem[\protect\citeauthoryear{Tiryaki and Kaplano{\u{g}}lu}{2022}]{tiryaki2022deep}
\begin{barticle}
\bauthor{\bsnm{Tiryaki}, \binits{V.}},
\bauthor{\bsnm{Kaplano{\u{g}}lu}, \binits{V.}}:
\batitle{Deep learning-based multi-label tissue segmentation and density assessment from mammograms}.
\bjtitle{IRBM}
\bvolume{43}(\bissue{6}),
\bfpage{538}--\blpage{548}
(\byear{2022})
\end{barticle}
\endbibitem

\bibitem[\protect\citeauthoryear{Dubrovina et~al.}{2018}]{dubrovina2018computational}
\begin{barticle}
\bauthor{\bsnm{Dubrovina}, \binits{A.}},
\bauthor{\bsnm{Kisilev}, \binits{P.}},
\bauthor{\bsnm{Ginsburg}, \binits{B.}},
\bauthor{\bsnm{Hashoul}, \binits{S.}},
\bauthor{\bsnm{Kimmel}, \binits{R.}}:
\batitle{Computational mammography using deep neural networks}.
\bjtitle{Computer Methods in Biomechanics and Biomedical Engineering: Imaging \& Visualization}
\bvolume{6}(\bissue{3}),
\bfpage{243}--\blpage{247}
(\byear{2018})
\end{barticle}
\endbibitem

\bibitem[\protect\citeauthoryear{Bou}{2019}]{bou2019deep}
\begin{botherref}
\oauthor{\bsnm{Bou}, \binits{A.}}:
Deep Learning models for semantic segmentation of mammography screenings
(2019)
\end{botherref}
\endbibitem

\bibitem[\protect\citeauthoryear{Nguyen et~al.}{2023}]{nguyen2023vindr}
\begin{barticle}
\bauthor{\bsnm{Nguyen}, \binits{H.T.}},
\bauthor{\bsnm{Nguyen}, \binits{H.Q.}},
\bauthor{\bsnm{Pham}, \binits{H.H.}},
\bauthor{\bsnm{Lam}, \binits{K.}},
\bauthor{\bsnm{Le}, \binits{L.T.}},
\bauthor{\bsnm{Dao}, \binits{M.}},
\bauthor{\bsnm{Vu}, \binits{V.}}:
\batitle{Vindr-mammo: A large-scale benchmark dataset for computer-aided diagnosis in full-field digital mammography}.
\bjtitle{Scientific Data}
\bvolume{10}(\bissue{1}),
\bfpage{277}
(\byear{2023})
\end{barticle}
\endbibitem

\bibitem[\protect\citeauthoryear{Zuiderveld}{1994}]{clahe}
\begin{bbook}
\bauthor{\bsnm{Zuiderveld}, \binits{K.}}:
\bbtitle{Contrast Limited Adaptive Histogram Equalization},
pp. \bfpage{474}--\blpage{485}.
\bpublisher{Academic Press Professional, Inc.},
\blocation{USA}
(\byear{1994})
\end{bbook}
\endbibitem

\bibitem[\protect\citeauthoryear{Sahin and Sahin}{2021}]{sahin2021introduction}
\begin{botherref}
\oauthor{\bsnm{Sahin}, \binits{{\"O}.}},
\oauthor{\bsnm{Sahin}, \binits{{\"O}.}}:
Introduction to apple ml tools.
Develop Intelligent iOS Apps with Swift: Understand Texts, Classify Sentiments, and Autodetect Answers in Text Using NLP,
17--39
(2021)
\end{botherref}
\endbibitem

\bibitem[\protect\citeauthoryear{Heath et~al.}{1998}]{heath1998current}
\begin{bchapter}
\bauthor{\bsnm{Heath}, \binits{M.}},
\bauthor{\bsnm{Bowyer}, \binits{K.}},
\bauthor{\bsnm{Kopans}, \binits{D.}},
\bauthor{\bsnm{Kegelmeyer~Jr}, \binits{P.}},
\bauthor{\bsnm{Moore}, \binits{R.}},
\bauthor{\bsnm{Chang}, \binits{K.}},
\bauthor{\bsnm{Munishkumaran}, \binits{S.}}:
\bctitle{Current status of the digital database for screening mammography}.
In: \bbtitle{Digital Mammography: Nijmegen, 1998},
pp. \bfpage{457}--\blpage{460}.
\bpublisher{Springer}, \blocation{???}
(\byear{1998})
\end{bchapter}
\endbibitem

\bibitem[\protect\citeauthoryear{Silva et~al.}{2023}]{silva2023data}
\begin{bchapter}
\bauthor{\bsnm{Silva}, \binits{S.V.}},
\bauthor{\bsnm{Sierra-Franco}, \binits{C.A.}},
\bauthor{\bsnm{Hurtado}, \binits{J.}},
\bauthor{\bsnm{Cruz}, \binits{L.C.}},
\bauthor{\bsnm{Thomaz}, \binits{V.d.A.}},
\bauthor{\bsnm{Silva-Calpa}, \binits{G.F.M.}},
\bauthor{\bsnm{Raposo}, \binits{A.B.}}:
\bctitle{A data-centric approach for pectoral muscle deep learning segmentation enhancements in mammography images}.
In: \bbtitle{International Symposium on Visual Computing},
pp. \bfpage{56}--\blpage{67}
(\byear{2023}).
\bcomment{Springer}
\end{bchapter}
\endbibitem

\end{thebibliography}
\fi


\end{document}